\documentclass[twocolumn,preprintnumbers,amssymb,amsmath,superscriptaddress,letterpaper,nofootinbib]{revtex4}[12pt]
\usepackage{times,graphics}
\usepackage[colorlinks]{hyperref}

\usepackage{graphicx}
\usepackage{dcolumn}
\usepackage{bm}
\usepackage{natbib}
\usepackage{pstricks}
\usepackage{epsfig}
\usepackage{epstopdf}
\usepackage{multirow}
\usepackage{amsmath}
\usepackage{lipsum}

\newcommand{\be}{\begin{equation}}
\newcommand{\ee}{\end{equation}}
\newcommand{\bea}{\begin{eqnarray}}
\newcommand{\eea}{\end{eqnarray}}

\makeatletter
\newcommand*{\shifttext}[2]{%
	\settowidth{\@tempdima}{#2}%
	\makebox[\@tempdima]{\hspace*{#1}#2}%
}
\makeatother

\begin{document}

\title{Ginzburg-Landau Theory of Dark Energy\\ A Framework to Study Both Temporal and Spatial Cosmological Tensions Simultaneously}
\author{Abdolali Banihashemi}
\email{a\_banihashemi@sbu.ac.ir}
\affiliation{Department of Physics, Shahid Beheshti University, G.C., Evin, Tehran 19839, Iran}

\author{Nima Khosravi}
\email{n-khosravi@sbu.ac.ir}
\affiliation{Department of Physics, Shahid Beheshti University, G.C., Evin, Tehran 19839, Iran}

\author{Amir H. Shirazi}
\email{amir.h.shirazi@gmail.com }
\affiliation{Department of Physics, Shahid Beheshti University, G.C., Evin, Tehran 19839, Iran}

\date{\today}

\begin{abstract}
A dark energy model (DE) is proposed based on Ginzburg-Landau theory of phase transition (GLT). This model, GLTofDE, surprisingly provides a framework to study not only temporal tensions in cosmology e.g. $H_0$ tension but also spatial anomalies of CMB e.g. the hemispherical power asymmetry and quadrupole-octopole alignment. In the mean field (or Landau) approximation of GLTofDE, there is a spontaneously symmetry breaking exactly like the Higgs potential. We modeled this transition, phenomenologically, and showed that GLTofDE can resolve both the $H_0$ tension and Lyman-$\alpha$ anomaly in a non-trivial way. According to $\chi^2$-analysis the transition happens at $z_t=0.738\pm0.028$ while $H_0=71.89\pm0.93$ km/s/Mpc and $\Omega^{like}_{k}=-0.225\pm0.049$ which are consistent with the latest $H(z)$ reconstructions. In addition, the GLTofDE proposes a framework to address the CMB anomalies when it is considered beyond the mean field approximation. In this regime existence of a long wavelength mode is a typical consequence which is named the Goldstone mode in the case of continuous symmetries. This mode, which is an automatic byproduct in GLTofDE, makes  cosmological constant, direction dependent. This means one side of the sky should be colder than the other side in agreement with what has been already observed in CMB. In addition between initial  stochastic pattern and the final state with one long wavelength mode, we can observe smaller patches or protrusions of the biggest remaining patch in the simulation. Our simulations show these protrusions are few in numbers and will be evolved according to Alan-Cahn mechanism. These protrusions can give an additional effect on CMB which is the existence of aligned quadrupole-octopole mode and its direction should be orthogonal to the dipole direction. We conclude that GLTofDE is a fertile framework both theoretically and phenomenologically.

\end{abstract}
\maketitle
\section{Introduction:}
The standard model of cosmology, $\Lambda$CDM including $\sim 68\%$ dark energy and $\sim 32\%$ dark matter and ordinary matter, can describe our universe very accurately thanks to precision cosmology era \cite{Aghanim:2018eyx}. Although $\Lambda$CDM is the best known model of our universe but it suffers from some issues in its details. Here we do not focus on theoretical problems like the cosmological constant \cite{CC-problem} or coincidence problem. But, we concentrate on the tensions between $\Lambda$CDM predictions and  parameters emerging from observations both in early time, e.g. CMB and late time, e.g. supernovae. The most important tension is in the prediction of the present value of the Hubble parameter, $H_0$. $\Lambda$CDM based on CMB data \cite{Aghanim:2018eyx} predicts a lower value for $H_0$ in comparison to that derived from direct measurements of supernovae \cite{riess18,Riess:2016jrr} with $\sim 4\sigma$ discrepancy. Another way to see this tension is in the behavior of $H(z)$ predicted by $\Lambda$CDM and that reconstructed from the data directly \cite{Bernal:2016gxb,Zhao:2017cud,Wang:2018fng,Dutta:2018vmq} where a $\sim 3.5 \sigma$ tension has been reported \cite{Zhao:2017cud}. At the level of the linear perturbation there is also a mild tension in $f\sigma_8$ which represents the matter content of the universe. Observations hint that there is less matter in the late time in comparison to what we expect from CMB \cite{Abbott:2017wau}. Another anomaly is in the distance measurement  of Lyman-$\alpha$ BAO \cite{Bautista:2017zgn,Bourboux:2017cbm}.  We categorize these tensions as temporal tensions since they suggest a difference between early and late time cosmologies. In addition there are spatial anomalies which have been reported in CMB anisotropies by both WMAP \cite{Bennett:2003bz} and Planck \cite{Ade:2015hxq} and reviewed in \cite{Schwarz:2015cma}. The hemisphere asymmetry stresses that temperature fluctuations have larger amplitude on one side of CMB than on another side. it means there is an unexpected dipole modulation in CMB power spectrum \cite{Eriksen:2003db}. Another famous anomaly is the presence of a cold spot in CMB \cite{Vielva:2003et}. The other anomaly is the alignment of quadrupole and octopole modes \cite{deOliveira-Costa:2003utu}. Note that these spatial anomalies are around $2-3\sigma$ which may not be so significant but the story is different if they are assumed to be independent\footnote{The covariance  of CMB anomalies has been studied in \cite{Muir:2018hjv}.} which means around $7\sigma$ tension in $\Lambda$CDM \cite{Hansen:2018pgg} and if one takes $H_0$ tension into the account then the overall tension becomes even more severe.  

These tensions are motivations for proposing many ideas in relevant literature. One candidate to address these tensions is considering the standard physics more carefully. For example the neutrinos always were a candidate to address these anomalies specially $H_0$ and $\sigma_8$ tensions \cite{Poulin:2018zxs}. Although it is an interesting idea inside the known physics but unfortunately it cannot solve the problems. In another viewpoint these tensions are hints of a new physics where the more interesting ideas are the ones which either have some roots in a fundamental (well-known) physics and/or solve more than one of these tensions at once. To address the temporal tensions specially $H_0$ tension, the physics of dark energy and gravity has been studied very extensively e.g.  interacting dark energy \cite{DiValentino:2017iww,Yang:2018euj}, neutrino-dark matter interaction \cite{DiValentino:2017oaw},  varying Newton constant \cite{Nesseris:2017vor}, viscous bulk cosmology \cite{Mostaghel:2016lcd}, phantom-like dark energy \cite{DiValentino:2016hlg}, early dark energy \cite{Poulin:2018cxd}, massive graviton \cite{DeFelice:2016ufg}  and etc. In addition modified gravity can address $f\sigma_8$ tension too \cite{DiValentino:2015bja}. It has been shown in \cite{Khosravi:2017hfi,DiValentino:2017rcr} that a transition in the behavior of gravitational force \cite{Khosravi:2017aqq} can lessen the $H_0$ tension. A transition in the bahavior of dark energy has been studied in \cite{Bassett:2002fe,Bassett:2002qu,Kobayashi:2018nzh} but not for solving the $H_0$ tensions.  Recently by focusing on the transition point we have shown that a DE model inspired by the Ising model can be a framework to think about the temporal tensions \cite{Banihashemi:2018oxo}. However to address the spatial tensions i.e. CMB anomalies, usually the physics of early universe has been modified from  its standard version. To address the hemisphere asymmetry an idea is the existence of a long wave mode \cite{Erickcek:2008jp} which has been studied very extensively. This mode cannot be realized in single field inflation as discussed in \cite{Erickcek:2008sm,Erickcek:2009at}. Another idea is to relate non-Gaussianity to the hemispherical asymmetry \cite{Schmidt:2012ky,Namjoo:2013fka}. In an interesting proposal based on isotropic non-Gaussian $g_{NL}$-like model,   almost all of these anomalies has been addressed simultaneously   in  \cite{Hansen:2018pgg}. Although usually for CMB anomalies the physics of inflation has been modified but there are few works based on  late time cosmology too e.g. \cite{Koivisto:2008ig}.

In this work we  propose a new model of dark energy based on the well-established physics of the critical phenomena. As we mentioned above, we have examined a simple model in \cite{Banihashemi:2018oxo} by assuming an Ising-inspired model for DE. Now we generalize our idea by assuming DE underwent a phase transition in its history in a model independent way. This idea can be realized by working within Ginzburg-Landau framework which is an effective field theory describing the physics of phase transition without any dependence on the details of relevant micro-structures. Since the physics of Ginzburg-Landau Theory (GLT) is very crucial for us in the next section \ref{GLT} we will focus on its details. In  section \ref{GLTDE} we will  propose a DE model based on GLT and will show how this model can address both temporal and spatial anomalies   of the standard cosmology, simultaneously, which happens for the first time in the literature up to our knowledge. We  finish with conclusions and future perspectives in the last section \ref{con}.

\section{Ginzburg-Landau Model}\label{GLT}
The main idea of GLT is to write an effective action for a macroscopic system by ignoring its microscopic details though keeping its main properties alive \cite{kardar}. This idea has been realized by writing a general phenomenological action for the critical phenomena. By coarse graining over the microscopic structure one can write an effective action as 
\begin{eqnarray}\label{GL-action}\nonumber
	{\cal H}_{GL}&=&C+\frac{1}{2}m^2\big(\frac{T-T_c}{T_c}\big)\Phi^2+\xi\Phi^3+\lambda\Phi^4+\gamma (\nabla\Phi)^2\\&+&\zeta\,H.\Phi+{\cal{O}}(\Phi,\nabla\Phi),
\end{eqnarray}
where $C$ is an overall constant\footnote{As it is mentioned in \cite{kardar} ``The integration over the magnetic and non-magnetic degrees of freedom at
short scales also generates an overall constant". This factor usually is ignored in critical phenomena analysis but we keep it for our cosmological model. Our physical intuition is that when gravity is at play then even an overall constant will gravitate which will see is the cosmological constant.}, $H$ is the external (magnetic) field and $\Phi$ is representing the coarse grained field. Note that  the symmetries tell which terms should be in ${\cal H}_{GL}$ exactly like what we expect from and effective theory. The first  non trivial and interesting property of the above action is the coefficient of $\Phi^2$ which changes its sign at the critical temperature $T_c$ resulting in a phase transition in the system. The other term which will be crucial in the above action is the gradient term, $(\nabla\Phi)^2$, which represents the interaction between neighboring cells in the lattice. We should mention that, in principle, one can add higher order terms to GL model, ${\cal{O}}(\Phi,\nabla\Phi)$, but they just change the quantitative analysis without any new qualitative behavior of the system so in this work we will work with the first few terms. In the following we will focus on properties of the above action which are well-known in the literature of critical phenomena \cite{rednerBook}.

\begin{figure}
	\centering
	\includegraphics[width=.9\linewidth]{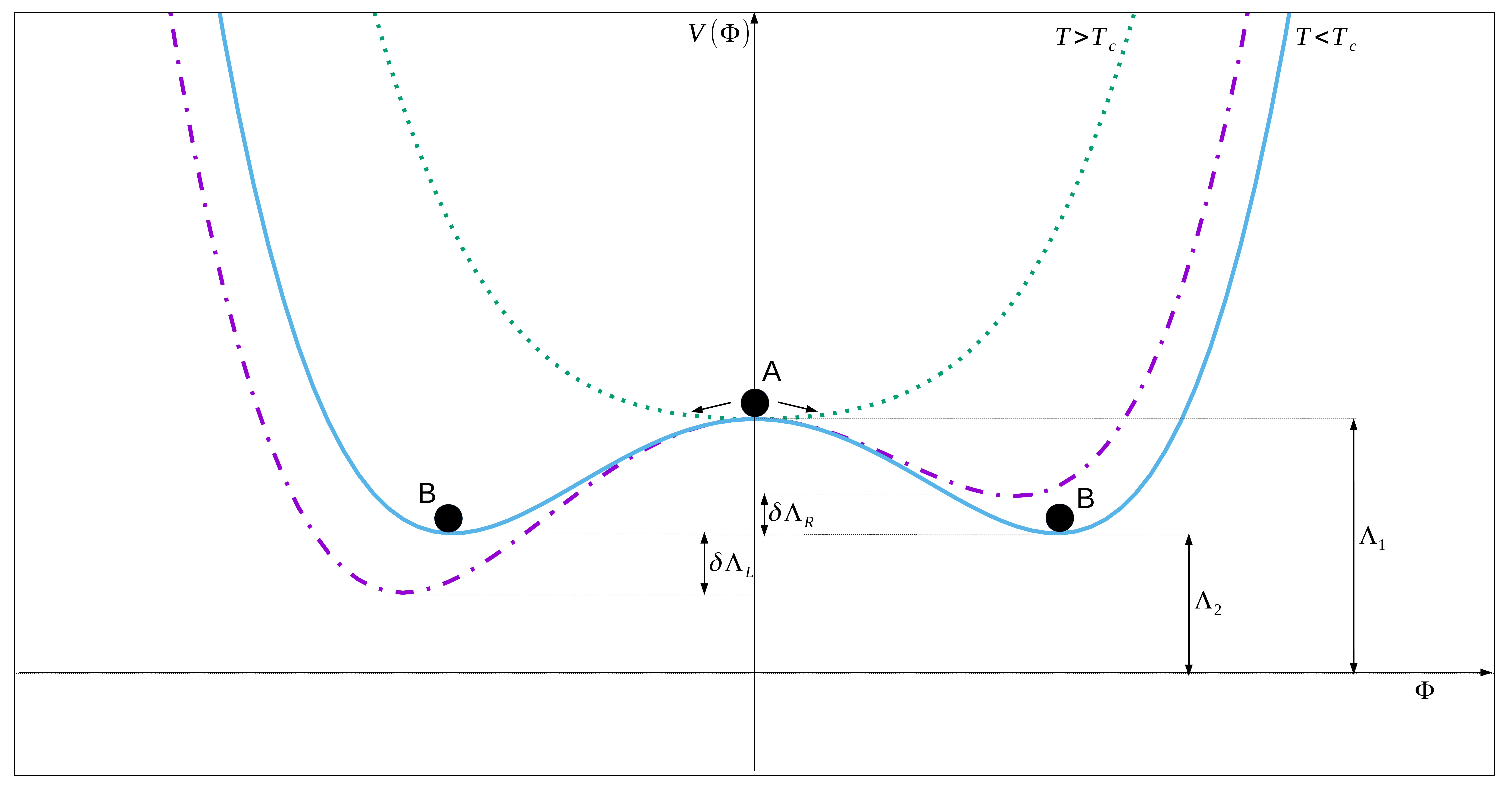}
	\caption{In this figure we have plotted GL-action for $T>T_c$ by dotted line and for $T<T_c$ by solid line. Before phase transition we do have $\Lambda_1$ as the value of potential which is bigger than its value after phase transition i.e. $\Lambda_2$. The existence of terms like $\Phi^3$ or $H.\Phi$ can break the $\mathbf{Z}_2$ symmetry and results in difference in the value of potential's minimum on the right and left. This difference will be crucial for us when we study the anisotropic part of our model.}
	\label{fig:potential}
\end{figure}

\subsection{Landau Approximation}
Landau made an approximation by assuming a coarse graining with a typical length comparable to the size of the lattice. This approximation is famous as mean field approximation and effectively says each cell can see all the cells in the lattice. In Landau approximation there is no spatial dependence in the field $\Phi$ and consequently ${\cal H}_{GL}$ will be\footnote{Without loss of generality in this section we ignore $\Phi^3$ and $H.\Phi$ terms.}
\begin{eqnarray}\label{Landau-action}
{\cal H}_{Landau}=\frac{1}{2}m^2\big(\frac{T-T_c}{T_c}\big)\Phi^2+\lambda\Phi^4,
\end{eqnarray}
which has been plotted in FIG. \ref{fig:potential} before and after $T_c$. Obviously the minimum field value, $\Phi_0$,  depends on the signature of $T-T_c$ as
\begin{eqnarray} \label{field-Landau}
\Phi_0 =
\begin{cases}
0 \qquad& T>T_c, \\
\pm\bigg(\frac{m^2(T_c-T)}{4\lambda T_c}\bigg)^{1/2} \qquad& T<T_c
\end{cases}
\end{eqnarray}
where gives the following values for the potential
\begin{eqnarray} \label{potential-Landau}
V(\Phi_0) =
\begin{cases}
 0\qquad& T>T_c, \\
-\frac{1}{16}\frac{m^4(T-T_c)^2}{\lambda T_c^2} \qquad& T<T_c
\end{cases}
\end{eqnarray}
with a lower value for potential after critical temperature i.e. $V_{bc} < V_{ac}$ as it is obvious in FIG. \ref{fig:potential}. In the case of magnetization it means above $T_c$ fluctuations are dominant and the net magnetization is zero. However by decreasing the temperature and below $T_c$, spins become aligned in macroscopic patches and there is a non vanishing net magnetization. For $T<T_c$, if we give enough time to the system then all the spins become aligned which is the final equilibrium state. However before reaching to this state, there is distribution of patches in the system. To study the effects of these patches we need to go beyond mean field approximation.

\subsection{Beyond Landau Approximation}
As we could see in Landau approximation, at $T=T_c$ a phase transition happens which causes a spontaneous symmetry breaking. For example in the case of two-dimensional $\Phi$ after the phase transition we have a Mexican hat potential and due to the (rotational) symmetry of the model there is no energy difference between the local minima. As we mentioned all the spins will be aligned at absolute zero but there is no preferred direction as a direct consequence of (rotational) symmetry \cite{kardar}. Note that this property is held even for all the temperature below $T_c$. Note that any spontaneous (continuous) symmetry breaking produces a long-wave mode which is famous as Goldstone mode. This mode has a long wavelength which cannot be explained by Landau approximation and it needs us to consider the last terms in (\ref{GL-action}), $(\nabla\Phi)^2$. This term means neighbor cells in the lattice has interaction with each other in a way to minimize the energy spins in the neighbor cells  to make them be aligned. This explains why the Goldstone mode has a long wavelength.

To have an idea about the behavior of GL model we have simulated its time dependent version i.e. ``Time Dependent Ginzburg Landau" (TDGL). For details one can see the Appendix \ref{TDGL} but the summary is in  or high temperatures we do have very small stochastically distributed patches and the lattice does not exhibit any longe range order. While below critical temperature one state will be dominated and correlation length diverges. Before the final state we have one big patch, where its size is comparable with the lattice size, which is exactly what we expected as the long-wavelength  mode. However more interestingly we can see before this state we have few smaller patches. These patches either are islands which will be swallowed by the sea or are  protrusions of the biggest patch which will dissolved into the main part. Note that these smaller patches cannot be very close to the biggest patch since they will quickly be dissolved. In FIG. \ref{fig:cartoon1}, these properties became visualized in a cartoon but details are given in Appendix \ref{TDGL}.

\section{Ginzburg-Landau Theory of Dark Energy}\label{GLTDE}
In this section we will propose an idea on dark energy based on GLT. We assume dark energy has a kind of microscopic structure which could undergo a phase transition. So effectively its Lagrangian can be given by Ginzburg-Landau model (\ref{GL-action}) without any concerns about the details of its microscopic structure. On the other hand in this work, for simplicity, we assume dark energy sector has a very small interaction with other species of the universe including the gravity sector. This approximation allows us to study the interesting and important properties of our model without loss of generality. It means we can solve TDGL equation to get dynamics of DE in our model and then plug the solution into the equations of motion of other species of the universe. So, we  assume the dynamics is governed by TDGL equation (\ref{equTDGL}) while the potential for DE is given as
\begin{eqnarray}\label{DE-action}\nonumber
V_{GLTofDE}&=&\Lambda+\frac{1}{2}m^2\big(\frac{T-T_c}{T_c}\big)\phi^2+\xi\phi^3+\lambda\phi^4\\&+&\zeta \rho_{ext} \phi+\gamma \nabla\phi.\nabla\phi,
\end{eqnarray}
where we introduced $\Lambda$ instead of $C$ in (\ref{GL-action}), $\phi$ is a scalar field and $\rho_{ext}$ represents any field except DE very similar to interacting DE models \cite{Farrar:2003uw,Amendola:2006dg}. The above potential should remind us the effective theory of dark energy which has been studied in the literature extensively \cite{Gubitosi:2012hu,Bellini:2014fua}. Although, the main conceptual difference  is the assumption of micro-structure for DE which results in $T-T_c$ factor in the second term practically. For our cosmological purposes we think $T$ is proportional to photon temperature and phase transition happens at  a transition redshift $z_c$ which corresponds to a critical temperature $T_c$ i.e. $(z-z_c)\propto(T-T_c)$.

\subsection{Background: Homogeneous and Isotropic Part}
It is worth to mention that the homogeneous and isotropic background part is exactly same as Landau model where the field has no spatial dependence. It is obvious why it is the case if we recall that the homogeneity and isotropy is an approximation for above $100\,\rm Mpc$ scales. Averaging over this scale is similar to coarse graining in Landau approximation. According to the field value transition in Landau model (\ref{field-Landau}) the amplitude of the potential will switch at the critical temperature from a higher value in higher redshifts to a lower one in lower redshifts\footnote{For our purpose in the background we assume $\xi\ll 1$ which is in agreement with non-Gaussianity observations by Planck results.}, see FIG. \ref{fig:potential}.
The $z$-dependence of this transition depends on the details of cooling procedure\footnote{There are two main cooling procedure i.e. annealing and quenching which are very slow and very fast respectively.} and we model it  by $\Lambda_{eff} =\Lambda\,X(z)$ with
\begin{eqnarray} \label{X-tanh}
X(z)=1+A\,\bigg[\tanh\big(\alpha(z-z_c)\big)+\tanh\big(\alpha\,z_c\big)\bigg]
\end{eqnarray}
where $z_c$ is representing the critical redshift, $A$ and $\alpha$ are the amplitude and the shape of the transition. Consequently, Hubble parameter for a homogeneous and isotropic universe will be modified as
\begin{eqnarray} \label{hubble}
&&H^2(z)=\\\nonumber &&H_0^2\big[\Omega_r\,(1+z)^4+\Omega_m\,(1+z)^3+\Omega^{like}_k\,(1+z)^2+\Omega_{\Lambda}\,X(z)\big]
\end{eqnarray}
where we assumed $\Omega_r+\Omega_m+\Omega^{like}_k+\Omega_{\Lambda}=1$ at $z=0$ which is consistent with definition of $X(z)$ and $H_0$ is Hubble parameter at $z=0$. Note that we introduced $\Omega^{like}_k$ term which is very similar to spatial curvature term but it is different since it just appears in the dynamics and not in the geometrical kinematics i.e. it means our model is spatially flat.

\begin{table}
	\begin{tabular}{ |c|c|c| }
		\hline
	Data sets	& $\Lambda$CDM & GLT of DE \\ \hline
		& &\\
		\multirow{9}{*}{\rotatebox[origin=c]{90}{Background }} & $\chi^2=24.80\qquad\gamma=1.77$ & $\chi^2=12.71\qquad\gamma=1.06$ \\
		&  &  \\
		& $H_0=71.13\pm0.80$ & $H_0=71.89\pm0.93$ \\
		&  &  \\
		& $\Omega_m h^2=0.1433\pm0.0064$ & $\Omega_m h^2=0.1432\pm0.0074$ \\
		&  &  \\
		& $\Omega^{like}_{k}=-0.046\pm0.012$ &  $\Omega^{like}_{k}=-0.225\pm0.049$\\
		&  &  \\
		&  & $A=0.80\pm0.19$ \\ 
		&  &  \\
		&  & $z_t=0.738\pm0.028$ \\
		& & \\ \hline
		&  &\\
		\multirow{5}{*}{\rotatebox[origin=c]{90}{ \hspace{-2cm}Background+$f\sigma_8(z)$  }} &$\chi^2=31.73\qquad\gamma=1.38$  & $\chi^2=20.33\qquad\gamma=0.97$ \\
			&  &  \\
		& $H_0=71.22\pm0.36$ & $H_0=71.82\pm0.91$ \\
		&  &  \\
		& $\Omega_m h^2=0.1431^{+0.0064}_{-0.0069}$ & $\Omega_m h^2=0.1431\pm0.0073$ \\
		&  &  \\
		& $\Omega^{like}_{k}=-0.045^{+0.011}_{-0.013}$ &  $\Omega^{like}_{k}=-0.204^{+0.045}_{-0.038}$\\
		&  &  \\
		&  & $A=0.72\pm0.17$ \\ 
		&  &  \\
		&  & $z_t=0.733\pm0.029$ \\
		& & \\ \hline
	\end{tabular}
	\caption{In this Table we have reported our $\chi^2$ analysis for two sets of the data points. Once we have used just background data points and then we have added $f\sigma_8(z)$ data points where the details can be found in Appendix \ref{data-sets-app}. Obviously our model is much better in $\chi^2$ analysis and can solve $H_0$ tension. However we do have more free parameters but it is also obvious that our model has much better reduced-$\chi^2$; which is defined as $\gamma=\chi^2_{min}/(N_{data}-N_{model})$, where $N_{data}$ is number of data points and $N_{model}$ is number of free parameters in the model.  An interesting property of GLTofDE is prediction of $z_t\sim 0.75$ which is consistent with the observations from $H(z)$ reconstruction. It is worth to mention that a negative value for $\Omega^{like}_k$ is consistent with the recent results in $H(z)$ reconstruction \cite{Wang:2018fng,Dutta:2018vmq} as we described in the Appendix \ref{data-sets-app}.}.\label{table:best-fit}
\end{table}

\begin{figure}
	\centering
\begingroup
\makeatletter
\providecommand\color[2][]{%
	\GenericError{(gnuplot) \space\space\space\@spaces}{%
		Package color not loaded in conjunction with
		terminal option `colourtext'%
	}{See the gnuplot documentation for explanation.%
	}{Either use 'blacktext' in gnuplot or load the package
		color.sty in LaTeX.}%
	\renewcommand\color[2][]{}%
}%
\providecommand\includegraphics[2][]{%
	\GenericError{(gnuplot) \space\space\space\@spaces}{%
		Package graphicx or graphics not loaded%
	}{See the gnuplot documentation for explanation.%
	}{The gnuplot epslatex terminal needs graphicx.sty or graphics.sty.}%
	\renewcommand\includegraphics[2][]{}%
}%
\providecommand\rotatebox[2]{#2}%
\@ifundefined{ifGPcolor}{%
	\newif\ifGPcolor
	\GPcolortrue
}{}%
\@ifundefined{ifGPblacktext}{%
	\newif\ifGPblacktext
	\GPblacktexttrue
}{}%
\let\gplgaddtomacro\g@addto@macro
\gdef\gplbacktext{}%
\gdef\gplfronttext{}%
\makeatother
\ifGPblacktext
\def\colorrgb#1{}%
\def\colorgray#1{}%
\else
\ifGPcolor
\def\colorrgb#1{\color[rgb]{#1}}%
\def\colorgray#1{\color[gray]{#1}}%
\expandafter\def\csname LTw\endcsname{\color{white}}%
\expandafter\def\csname LTb\endcsname{\color{black}}%
\expandafter\def\csname LTa\endcsname{\color{black}}%
\expandafter\def\csname LT0\endcsname{\color[rgb]{1,0,0}}%
\expandafter\def\csname LT1\endcsname{\color[rgb]{0,1,0}}%
\expandafter\def\csname LT2\endcsname{\color[rgb]{0,0,1}}%
\expandafter\def\csname LT3\endcsname{\color[rgb]{1,0,1}}%
\expandafter\def\csname LT4\endcsname{\color[rgb]{0,1,1}}%
\expandafter\def\csname LT5\endcsname{\color[rgb]{1,1,0}}%
\expandafter\def\csname LT6\endcsname{\color[rgb]{0,0,0}}%
\expandafter\def\csname LT7\endcsname{\color[rgb]{1,0.3,0}}%
\expandafter\def\csname LT8\endcsname{\color[rgb]{0.5,0.5,0.5}}%
\else
\def\colorrgb#1{\color{black}}%
\def\colorgray#1{\color[gray]{#1}}%
\expandafter\def\csname LTw\endcsname{\color{white}}%
\expandafter\def\csname LTb\endcsname{\color{black}}%
\expandafter\def\csname LTa\endcsname{\color{black}}%
\expandafter\def\csname LT0\endcsname{\color{black}}%
\expandafter\def\csname LT1\endcsname{\color{black}}%
\expandafter\def\csname LT2\endcsname{\color{black}}%
\expandafter\def\csname LT3\endcsname{\color{black}}%
\expandafter\def\csname LT4\endcsname{\color{black}}%
\expandafter\def\csname LT5\endcsname{\color{black}}%
\expandafter\def\csname LT6\endcsname{\color{black}}%
\expandafter\def\csname LT7\endcsname{\color{black}}%
\expandafter\def\csname LT8\endcsname{\color{black}}%
\fi
\fi
\setlength{\unitlength}{0.0500bp}%
\ifx\gptboxheight\undefined%
\newlength{\gptboxheight}%
\newlength{\gptboxwidth}%
\newsavebox{\gptboxtext}%
\fi%
\setlength{\fboxrule}{0.5pt}%
\setlength{\fboxsep}{1pt}%
\begin{picture}(5100.00,3400.00)%
\gplgaddtomacro\gplbacktext{%
	\csname LTb\endcsname
	\put(362,396){\makebox(0,0)[r]{\strut{}$55$}}%
	\csname LTb\endcsname
	\put(362,972){\makebox(0,0)[r]{\strut{}$60$}}%
	\csname LTb\endcsname
	\put(362,1548){\makebox(0,0)[r]{\strut{}$65$}}%
	\csname LTb\endcsname
	\put(362,2123){\makebox(0,0)[r]{\strut{}$70$}}%
	\csname LTb\endcsname
	\put(362,2699){\makebox(0,0)[r]{\strut{}$75$}}%
	\csname LTb\endcsname
	\put(362,3275){\makebox(0,0)[r]{\strut{}$80$}}%
	\csname LTb\endcsname
	\put(430,272){\makebox(0,0){\strut{}$0$}}%
	\csname LTb\endcsname
	\put(1289,272){\makebox(0,0){\strut{}$0.5$}}%
	\csname LTb\endcsname
	\put(2147,272){\makebox(0,0){\strut{}$1$}}%
	\csname LTb\endcsname
	\put(3006,272){\makebox(0,0){\strut{}$1.5$}}%
	\csname LTb\endcsname
	\put(3865,272){\makebox(0,0){\strut{}$2$}}%
	\csname LTb\endcsname
	\put(4723,272){\makebox(0,0){\strut{}$2.5$}}%
}%
\gplgaddtomacro\gplfronttext{%
	\csname LTb\endcsname
	\put(-020,1835){\rotatebox{-270}{\makebox(0,0){\strut{}$H(z)/(1+z)[\rm kms^{-1}Mpc^{-1}]$}}}%
	\csname LTb\endcsname
	\put(2662,86){\makebox(0,0){\strut{}$z$}}%
	\begin{scriptsize}
	\csname LTb\endcsname
	\put(4370,3164){\makebox(0,0)[r]{\strut{}GLT background and $f\sigma_8$}}%
	\csname LTb\endcsname
	\put(4370,3040){\makebox(0,0)[r]{\strut{}$\Lambda$CDM background and $f\sigma_8$}}%
	\csname LTb\endcsname
	\put(4370,2916){\makebox(0,0)[r]{\strut{}GLT background}}%
	\csname LTb\endcsname
	\put(4370,2792){\makebox(0,0)[r]{\strut{}$\Lambda$CDM background}}%
	\csname LTb\endcsname
	\put(4370,2668){\makebox(0,0)[r]{\strut{}Planck}}%
	
	\end{scriptsize}
	\begin{footnotesize}
	\csname LTb\endcsname
	\put(1200,2488){\makebox(0,0)[r]{\strut{}Riess et al.}}%
	\csname LTb\endcsname
	\put(1900,588){\makebox(0,0)[r]{\strut{}BOSS DR12}}%
	\csname LTb\endcsname
	\put(3800,600){\makebox(0,0)[r]{\strut{}DR14 quasars}}%
	\csname LTb\endcsname
	\put(4750,1390){\makebox(0,0)[r]{\strut{}BOSS Ly-$\alpha$}}%
	\end{footnotesize}
}%
\gplbacktext
\put(0,0){\includegraphics{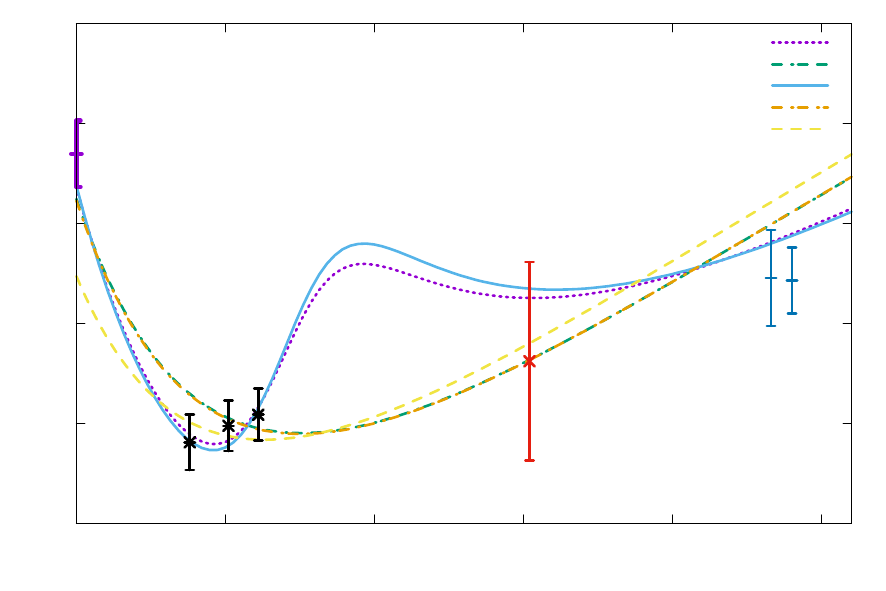}}%
\gplfronttext
\end{picture}%
\endgroup

	\caption{We have plotted $H(z)$ vs. redshift for GLTofDE and $\Lambda$CDM. $\Lambda$CDM has plotted for Planck 2018 best fit, with our background data points and background $+ f\sigma_8(z)$ data points in dashed, yellow and green dot-line respectively. $GLTofDE$ has been plotted in solid and dotted lines for different sets of data points. Obviously $GLTofDE$ has this potential to solve both $H_0$  and Lyman-$\alpha$ tensions simultaneously. One more interesting property is the behavior of $GLTofDE$ which can go through BOSS-DR12 data points properly. Obviously $GLTofDE$ has its own fingerprint e.g. its prediction for $H(z)$ around $z\sim 1$ is totally different with $\Lambda$CDM.}
	\label{fig:Hz}
\end{figure}

\begin{figure}
	\centering
	\begingroup
	\makeatletter
	\providecommand\color[2][]{%
		\GenericError{(gnuplot) \space\space\space\@spaces}{%
			Package color not loaded in conjunction with
			terminal option `colourtext'%
		}{See the gnuplot documentation for explanation.%
		}{Either use 'blacktext' in gnuplot or load the package
			color.sty in LaTeX.}%
		\renewcommand\color[2][]{}%
	}%
	\providecommand\includegraphics[2][]{%
		\GenericError{(gnuplot) \space\space\space\@spaces}{%
			Package graphicx or graphics not loaded%
		}{See the gnuplot documentation for explanation.%
		}{The gnuplot epslatex terminal needs graphicx.sty or graphics.sty.}%
		\renewcommand\includegraphics[2][]{}%
	}%
	\providecommand\rotatebox[2]{#2}%
	\@ifundefined{ifGPcolor}{%
		\newif\ifGPcolor
		\GPcolortrue
	}{}%
	\@ifundefined{ifGPblacktext}{%
		\newif\ifGPblacktext
		\GPblacktexttrue
	}{}%
	\let\gplgaddtomacro\g@addto@macro
	\gdef\gplbacktext{}%
	\gdef\gplfronttext{}%
	\makeatother
	\ifGPblacktext
	\def\colorrgb#1{}%
	\def\colorgray#1{}%
	\else
	\ifGPcolor
	\def\colorrgb#1{\color[rgb]{#1}}%
	\def\colorgray#1{\color[gray]{#1}}%
	\expandafter\def\csname LTw\endcsname{\color{white}}%
	\expandafter\def\csname LTb\endcsname{\color{black}}%
	\expandafter\def\csname LTa\endcsname{\color{black}}%
	\expandafter\def\csname LT0\endcsname{\color[rgb]{1,0,0}}%
	\expandafter\def\csname LT1\endcsname{\color[rgb]{0,1,0}}%
	\expandafter\def\csname LT2\endcsname{\color[rgb]{0,0,1}}%
	\expandafter\def\csname LT3\endcsname{\color[rgb]{1,0,1}}%
	\expandafter\def\csname LT4\endcsname{\color[rgb]{0,1,1}}%
	\expandafter\def\csname LT5\endcsname{\color[rgb]{1,1,0}}%
	\expandafter\def\csname LT6\endcsname{\color[rgb]{0,0,0}}%
	\expandafter\def\csname LT7\endcsname{\color[rgb]{1,0.3,0}}%
	\expandafter\def\csname LT8\endcsname{\color[rgb]{0.5,0.5,0.5}}%
	\else
	\def\colorrgb#1{\color{black}}%
	\def\colorgray#1{\color[gray]{#1}}%
	\expandafter\def\csname LTw\endcsname{\color{white}}%
	\expandafter\def\csname LTb\endcsname{\color{black}}%
	\expandafter\def\csname LTa\endcsname{\color{black}}%
	\expandafter\def\csname LT0\endcsname{\color{black}}%
	\expandafter\def\csname LT1\endcsname{\color{black}}%
	\expandafter\def\csname LT2\endcsname{\color{black}}%
	\expandafter\def\csname LT3\endcsname{\color{black}}%
	\expandafter\def\csname LT4\endcsname{\color{black}}%
	\expandafter\def\csname LT5\endcsname{\color{black}}%
	\expandafter\def\csname LT6\endcsname{\color{black}}%
	\expandafter\def\csname LT7\endcsname{\color{black}}%
	\expandafter\def\csname LT8\endcsname{\color{black}}%
	\fi
	\fi
	\setlength{\unitlength}{0.0500bp}%
	\ifx\gptboxheight\undefined%
	\newlength{\gptboxheight}%
	\newlength{\gptboxwidth}%
	\newsavebox{\gptboxtext}%
	\fi%
	\setlength{\fboxrule}{0.5pt}%
	\setlength{\fboxsep}{1pt}%
	\begin{picture}(5100.00,3400.00)%
	\gplgaddtomacro\gplbacktext{%
		\csname LTb\endcsname
		\put(747,909){\makebox(0,0)[r]{\strut{}$0.9$}}%
		\csname LTb\endcsname
		\put(747,1433){\makebox(0,0)[r]{\strut{}$0.95$}}%
		\csname LTb\endcsname
		\put(747,1956){\makebox(0,0)[r]{\strut{}$1$}}%
		\csname LTb\endcsname
		\put(747,2480){\makebox(0,0)[r]{\strut{}$1.05$}}%
		\csname LTb\endcsname
		\put(747,3004){\makebox(0,0)[r]{\strut{}$1.1$}}%
		\csname LTb\endcsname
		\put(849,409){\makebox(0,0){\strut{}$0$}}%
		\csname LTb\endcsname
		\put(1607,409){\makebox(0,0){\strut{}$0.5$}}%
		\csname LTb\endcsname
		\put(2366,409){\makebox(0,0){\strut{}$1$}}%
		\csname LTb\endcsname
		\put(3124,409){\makebox(0,0){\strut{}$1.5$}}%
		\csname LTb\endcsname
		\put(3883,409){\makebox(0,0){\strut{}$2$}}%
		\csname LTb\endcsname
		\put(4641,409){\makebox(0,0){\strut{}$2.5$}}%
	}%
	\gplgaddtomacro\gplfronttext{%
		\csname LTb\endcsname
		\put(153,1904){\rotatebox{-270}{\makebox(0,0){\strut{}$D_V/D_{V_{\rm Planck}}$}}}%
		\csname LTb\endcsname
		\put(2821,130){\makebox(0,0){\strut{}$z$}}%
		\csname LTb\endcsname
		\put(4005,3046){\makebox(0,0)[r]{\strut{}$D_V$}}%
		\csname LTb\endcsname
		\put(4005,2860){\makebox(0,0)[r]{\strut{}$D_M$}}%
		\begin{tiny}
		\csname LTb\endcsname
		\put(950,1920){\rotatebox{-270}{\makebox(0,0)[r]{\strut{}6DFGS}}}%
		\end{tiny}
		\begin{footnotesize}
		\csname LTb\endcsname
		\put(1350,2820){\makebox(0,0)[r]{\strut{}MGS}}%
		\csname LTb\endcsname
		\put(2100,2820){\makebox(0,0)[r]{\strut{}WiggleZ}}%
		\csname LTb\endcsname
		\put(3800,2480){\makebox(0,0)[r]{\strut{}SDSS quasars}}%
		\csname LTb\endcsname
		\put(4450,1000){\makebox(0,0)[r]{\strut{}Ly-$\alpha$($D_M$)}}%
		\csname LTb\endcsname
		\put(2920,1760){\makebox(0,0)[r]{\strut{}DES($D_M$)}}%
		\csname LTb\endcsname
		\put(2520,1060){\makebox(0,0)[r]{\strut{}DR14 LRG}}%
		\end{footnotesize}
		\begin{tiny}
		\csname LTb\endcsname
		\put(1360,1930){\rotatebox{-270}{\makebox(0,0)[r]{\strut{}DR12}}}%
		\end{tiny}

	}%
	\gplbacktext
	\put(0,0){\includegraphics{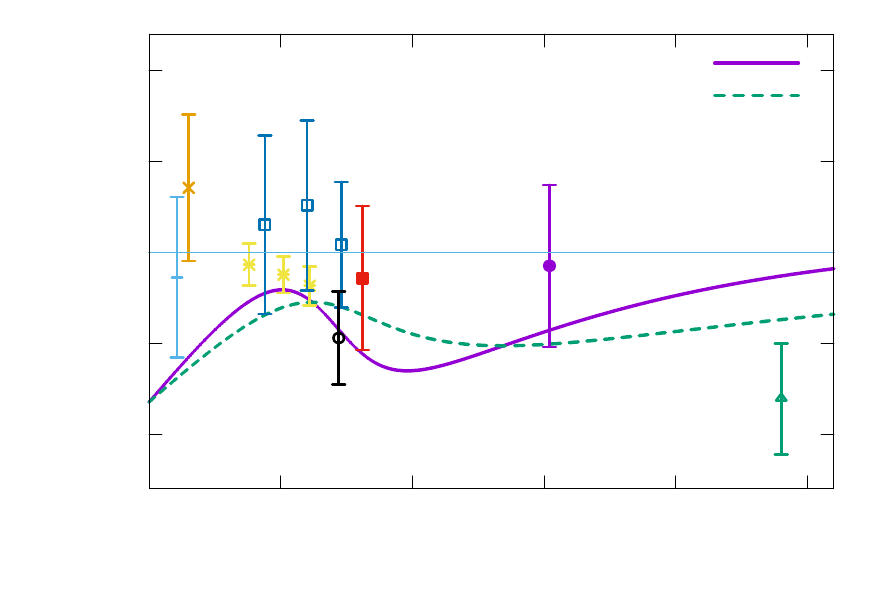}}%
	\gplfronttext
	\end{picture}%
	\endgroup
	
	\caption{We have plotted normalized $D_V(z)$ and $D_M(z)$ for GLTofDE in solid and dashed lines respectively. Note that we have not used some of these data points in our $\chi^2$ analysis but GLTofDE is very consistent with distance data points. Interestingly, GLTofDE predicted a very non-trivial behavior of $D_V(z)$ in $z\sim 0.4-0.7$ which follows the trend of BOSS-DR12 and DR14-LRG data points. Obviously GLTofDE (almost) solve  Lyman-$\alpha$ tension by predicting less $D_M(z)$ around $z\sim 2.4$ while it is compatible with  DES data point around $z\sim 0.7$.}
	\label{fig:DV}
\end{figure}

We have checked GLTofDE with background data encoded in $H(z)$ where we report their details in Appendix \ref{data-sets-app}. The best fits are reported in TABLE \ref{table:best-fit} and the details of likelihoods in FIG. \ref{fig:contour-back} in the Appendix \ref{data-sets-app}. In our analysis we have not run MCMC for parameter $\alpha$ which is the speed of phase transition in (\ref{X-tanh}) and we just work with a typical value of $\alpha=5.0$. This is because we could see our model is not too much sensitive to the value of $\alpha$ and on the other hand for large values of $\alpha$, $tanh$-functions behave like a step function and we cannot distinguish large $\alpha$'s at all. In addition to $H(z)$ data set we have added $f\sigma_8(z)$ data points to constrain our model (The details can be found in Appendix \ref{data-sets-app}.). Although we do not do perturbation theory of our model in this work but using $f\sigma_8(z)$ data points is still valid for our model. In GLTofDE as well as quintessence models the evolution equation of perturbations just be modified through the modifications in $H(z)$. So it is physically viable to use $f\sigma_8(z)$ data points in addition to $H(z)$'s. The best fits have been shown in  TABLE \ref{table:best-fit}. To get some intuition about the behavior of GLTofDE we have plotted $H(z)$ and $D_V(z)$ versus redshift in FIG. \ref{fig:Hz} and \ref{fig:DV} respectively. We can see that our model can perfectly describe all the data points with an obvious transition around $z_t\sim 0.75$. We would like to emphasize that our model has its very own fingerprints in  $H(z)$ and $D_V(z)$ between $z\sim 0.5-1.5$ which can be checked in future surveys like Euclid or SKA which look at structures in higher redshifts. One can find more details on the results in Appendix \ref{data-sets-app} including likelihoods in FIGS. \ref{fig:contour-back} and \ref{fig:contour-sigma}.

\begin{figure*}
	\centering
	\includegraphics[width=.9\linewidth]{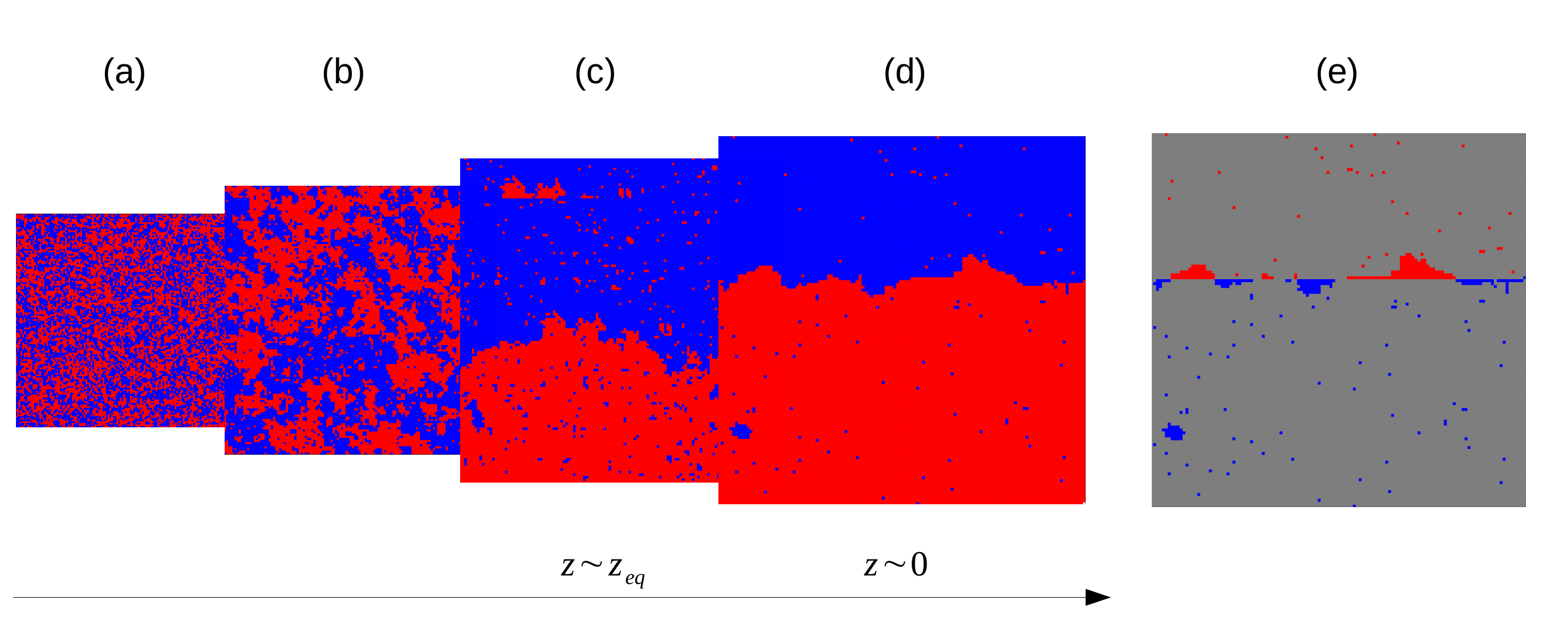}
	\caption{A cartoon sketched based on our simulations for $T<T_c$ (see Appendix \ref{TDGL}).  In very early times i.e. figure (a), the system is very stochastic but while time is going then we see the appearance of structures i.e. patches in figure (b). The final state will be as half red/blue (figures (c) and (d)) but then one color will be dominant which has not been in this figure. In cosmological scenario red and blue colors represent $\Lambda_2-\delta\Lambda_L$ and $\Lambda_2+\delta\Lambda_R$ (c.f. FIG. \ref{fig:potential}) which means different effective CC. Hence for redshifts before DM-DE equality i.e. $z>z_{eq}$, CC has no effect so we do not see any effects of patches in our model. But near $z_{eq}$ these patches will affect the cosmology and for our purposes we expect $z_{eq}$ be a little bit before (almost) final state i.e. figures (c) and (d). Obviously there is a dipole in (d) which can address the hemispherical asymmetry in CMB via (integrated) Sachs-Wolfe effect of different CC's. In figure (e) we removed the dipole structure and we can see the remaining gives a structure with higher multi-poles. In this simulation we can see three cold/hot patches which gives aligned quadrupole and octopole very interestingly. In addition in bottom-left of (e) we could get a cold/hot spot. We emphasize that for quantitative arguments we need more simulations which will be remained for the future works. However GLTofDE is a rich framework to address CMB anomalies as well as $H_0$ tension.} 
	\label{fig:cartoon1}
\end{figure*}

\subsection{Anisotropic Part}
Now we study the effects of anisotropic terms on the background i.e. beyond mean field approximation. From the previous section we expect to have patches in the sky with different values of cosmological constant, (CC)\footnote{Note that we assumed the field is at the minimum of the potential in FIG. \ref{fig:potential} which means effectively we have a constant cosmological constant in each patches.}. In this section for our purposes we keep $\phi^3$ term in (\ref{DE-action}). So the values of CC is given by $\Lambda_2+\delta\Lambda_R$ and $\Lambda_2-\delta\Lambda_L$ e.g. for blue and red patches\footnote{Note that for this purpose we could consider interaction term in (\ref{DE-action}) i.e. $\rho_{ext}\phi$ with the same effect as $\phi^3$. A term like $\rho_{ext}\phi$ will break the symmetry of black and white patches.} in FIG. \ref{fig:patch}. We emphasize that the existence of the patches as well as occurring the phase transition is because of the local interactions between neighborhoods given by $(\nabla\phi)^2$ term in (\ref{DE-action}) as we have discussed in the previous section. We remind that the existence of a long-wavelength mode is a natural consequence of broken symmetry.

In the cosmological setup it means we have a spatial asymmetry in the sky\footnote{In GLTofDE we should be careful about the DM-DE equality redshift, $z_{eq}$, because before $z_{eq}$ we do not expect to have any effects of DE including its patches.}. We have sketched a cartoon based on our simulations in FIG. \ref{fig:cartoon1}\footnote{It is actually based on a 2-D Ising simulation but the main concepts are same as what we have for 3-D GLT.}. This asymmetry can be seen in CMB due to Sachs-Wolfe effect\footnote{Integrated Sachs-Wolfe effects should be modified too but it is a secondary effect.}. I.e. since the late time acceleration depends on the direction, consequently, CMB photons will be affected by it.  The first biggest patch which remains till the end makes an obvious dipole in the sky according to different values of $\Lambda$'s for each side. The patch with larger $\Lambda$  causes larger redshift and consequently a colder side. We emphasize that this asymmetry is a direct consequence of having a phase transition in the dark energy behavior. This asymmetry in the sky can explain why a hemisphere  of the CMB sky is colder than the other side. For this purpose we need to assume the biggest patch (which is almost comparable with the Hubble radius) was dominant after DM-DE equality time\footnote{Even the biggest patch will be resolved if one gives enough time. Though its effect can be smaller but since CMB had seen this anisotropy in their history so this will be detectable even after reaching to the final state (i.e. the lattice with just one state, blue or red in FIG \ref{fig:cartoon1}.).}. So GLTofDE for sure predicts a dipole asymmetry in CMB. This result by itself is very interesting since it shows GLTofDE can address both $H_0$ and hemisphere asymmetry of CMB simultaneously.

Technically,  the power asymmetry is a dipole modulation, $\Theta^{obs}(\hat{n})=(1+A \hat{n}.\hat{p})\Theta^{iso}(\hat{n})$ \cite{Gordon:2006ag}, where $\Theta\equiv\Delta T/T$ and $\Theta^{obs}(\hat{n})$ and $\Theta^{iso}(\hat{n})$ are observed temperature and isotropic power respectively as functions of $\hat{n}$ i.e. the direction of observation. In this modulation $A$ is the amplitude of the anisotropic modulations and $\hat{p}$ is a preferred direction. An important observational property of this modulation is its appearance in only $\ell\le 60$ which makes it scale dependent. The most common idea to address this modulation as we mentioned already is the existence of a long-wave mode in the initial perturbations (i.e. during inflation) \cite{Erickcek:2008jp,Erickcek:2008sm,Erickcek:2009at}. In this framework the perturbations see this long-wave mode as a modulations over their isotropic background. In our GLTofDE scenario we do have the same long-wave mode modulations in the late time cosmology. As we mentioned above this late time long-wave mode will change the expansion rate in different directions of the sky and will cause a modulations in the power spectrum. We emphasize that this mode is time dependent since as times go up we expect that one of the patches become larger and larger and finally cover the whole sky. Note that the size of the smaller patch will makes the dipole modulation scale dependent. So its current size in the sky is constrained due to observation of the existence of the dipole modulations for $\ell\le 60$. However since in GLTofDE the smaller patch will be smaller and smaller and finally will be resolved completely then we predict that the dipole modulation will be observed in the smaller modes i.e. $\ell> 60$ but its amplitude depends on the patch's life time. But since the resolving speed is higher while the patch is smaller then we expect to have very small amplitudes for high $\ell$'s.

We can go further by considering the state of our model earlier than its final state. As we mentioned in Appendix \ref{TDGL}  Alan-Cahn mechanism says a patch will be evolved to reduce its curvature and make a circle/sphere and finally to be washed out. But before it became a symmetrical sphere we expect this patch can have non-symmetric protrusions like an octopus with non-symmetric arms. These arms are not symmetric in both size and position. But they cannot be very close to each other since their dynamics make them one arm. The Alan-Cahn mechanism wants to make this octopus a very symmetric octopus so we expect during this evolution there is a time that only a few (e.g. two or three) arms exist with different sizes. Now if this is the situation  after DM-DE equality then CMB should be affected by this structure. The main body (as we mentioned already) will make a dipole and the arms can produce both quadrupole and octopole which are automatically aligned \footnote{For completeness we would like to mention that in addition to protrusions we could think of smaller disjoint patches. Although our arguments  work for this scenario too, our simulations show that this scenario is less probable.}. In FIG. \ref{fig:cartoon1} we have sketched a cartoon describing these properties. This means GLTofDE framework can address two anomalies together; hemispherical power asymmetry and quadrupole-octopole alignment \footnote{We should check the details of statistic in details. This needs full consideration of simulations which is beyond the scope of this work and will remain for the future works.}. We also can imagine that the cold spot in CMB is the remnant of a disjoint patch which became like a sphere in its final state according to the Alan-Cahn mechanism. However to have this patch with appropriate size we need too much fine-tuning.

In this section we showed that ``GLT-of-DE" initiates  a very promising framework to study both temporal and spatial cosmological tensions  simultaneously. Up to our knowledge it makes our proposal unique in the literature.

\section{Concluding remarks and future perspective}\label{con}
We assumed (micro-)structure for dark energy which made a phase transition in its history. This phase transition can be realized by Ginzburg-Landau Theory which is the effective theory of phase transition. We could show GLTofDE can be a framework not only to address temporal tensions  of cosmology e.g. $H_0$ tension but also spatial ones i.e. CMB anomalies. We analyzed the background cosmology with backgrund data points (and $f\sigma_8(z)$ data) and our model is much better than $\Lambda$CDM in $\chi^2$ analysis. With our analysis the transition has been occurred around $z\sim 0.74$. Our model has its very own fingerprints between $z\sim 0.5-1.5$ in $H(z)$ and $D_V(z)$ as it is plotted in FIGS. \ref{fig:Hz} and \ref{fig:DV} which can be checked in near future surveys like Euclid and SKA. An unavoidable consequence of GLT is  the existence of a long-wavelength mode. This long-wavelength mode can affect the CMB temperature fluctuation and address the hemispherical asymmetry of CMB. As we have shown in FIG. \ref{fig:patch} this long-wavelength mode is the biggest patch of in the simulation. In addition the few smaller patches which could be survived after matter domination era can cause the alignment of quadrupole and octopole modes. We argued that quadrupole and octopole should be aligned and orthogonal to the dipole. This (natural) result of GLTofDE gives a promising framework to address CMB anomalies very interestingly. We also argue that the existence of the cold spot can be studied in this model but it needs fine-tuning. 

\subsection{Future perspective}
GLTofDE shows a promising framework with many directions to explore. The first direction is to work with CMB temperature/polarization data sets and employing Bayesian analysis to study goodness of fits. Another interesting direction is to investigate the CMB spatial anomalies more carefully. This needs to run very accurate simulations and study the details of results especially the distribution of patches' size vs. time. In this work we modeled the phase transition by a $tanh$-function but in principle one could try to solve the equations more concretely.  One further goal can be look for another observational fingerprints of our model e.g. between different patches we can expect to have domain walls and they should have their own affects.

At the end we would like to emphasize that GLTofDE  is based on a very profound and well-studied topic in physics i.e. critical phenomena. The idea of phase-transition can tell us more about our model GLTofDE e.g. in \cite{Eckel:2017uqx} it has been shown that a Bose condensate state can be seen as an expanding universe in the lab. In this direction we think it is possible to setup an experiment to simulate our idea. Actually it can also gives us more idea about the behavior of GLTofDE e.g. after the phase transition it is possible for the field to oscillate at the bottom of the potential before becoming relaxed to its final state\footnote{We thank A. Sheykhan and A. Ashoorioon for discussion on this point.} as it has been seen in \cite{Eckel:2017uqx}. This feature can explain the oscillations which has been observed in reconstructed $H(z)$ in \cite{Wang:2018fng}.

In addition we showed that GLTofDE is a very promising framework to think about the both temporal and spatial anomalies in the cosmology simultaneously which happens for the first time in the literature up to our knowledge.

\section*{Acknowledgments}
We are grateful to S. Baghram, M. Farhang, A. Mehrabi and S. Shahbazian for very fruitful discussions. We also thank A. Kargaran, A. Inanlou, A. Manavi and M. Sarikhani for their helps on using packages. NK would like to thank School of Physics at IPM where he is a resident researcher. We have used GetDist package to produce likelihoods \cite{getdist}.

\onecolumngrid

\appendix
\section{Time Dependent Ginzburg-Landau}\label{TDGL}
Using energy function, someone can go further and introduce a thermodynamic force $-\delta {\cal H} / \delta \Phi$ to write a dynamic equation in first order (overdamped) approximation:
\begin{eqnarray} \label{equTDGL}
\frac{\partial \Phi}{\partial t} &=& - \frac{\delta {\cal H}}{\delta \Phi} =\\\nonumber&& \nabla^2 \Phi + \frac{1}{2} m \big(\frac{T-T_c}{T_c}\big) \Phi +\xi \Phi^2+ \lambda \Phi^3 + \eta(t)
\end{eqnarray}
which is called \textit{Time-Dependent Ginzburg-Landau} (TDGL) equation. Last term $\eta(t)$ in TDGL represents thermal noise contribution, but many efforts of TDGL analysis are focused on zero-temperature dynamics. TDGL equation is the continuous counterpart of Ising-Glauber mechanism to simulate the relaxation process of a spin system after quenching to a zero temperature. FIG. \ref{figPlaneSample} demonstrates some steps of TDGL simulation, starting from a random state and ending to a global minimum. The most prominent aspect of TDGL analysis is interface evolution. Because of bilateral stable potential term, interfaces emerges after a few steps. \textit{Allen-Cahn equation} is governing the evolution of interfaces as follow \cite{rednerBook}:
\begin{align}\label{equAllenCahn}
v = - \sum_j^{d-1} \frac{1}{R_j}
\end{align}
where $v$ is the local velocity of an interface and $R_j$'s are principal radii of curvatures. So the fate of a $3d$ spherical droplet with radius $R_0$ is vanishing after $t = R_0^2 / 4$. Regarding this argument, final state in three dimension is one dominant state as ground state. The stability of an interface between two states only exist where all part of an interface has two opposite curvature states, i.e. saddle point. These interfaces are called minimal surfaces an only are presented in period boundary condition \cite{rednerBook}. Another important consequence of \textit{Allen-Cahn} is smoothing effect of TDGL dynamics. 

Based on TDGL evolution, we can study the evolution of discrete patches in three dimension. The simulation detail is as follow: we start with a random state in $100^3$ cubic lattice, and solve step-wise TDGL partial differential equation:
\begin{align} \label{equTDGLSim}
\Phi(x, t+dt) = \Phi(x, t) + dt \left( \nabla^2 \Phi + 2 \Phi \left(1-\Phi^2 \right) \right)
\end{align}
which is simplified version of (\ref{equTDGL}). One of main feature for our further argument is the statistics of $2d$ patches on plane intersections in the cubic lattice. For several time $t$, we sample $100$ different plane cross-section with cubic lattice and measure the size of discrete patches of $\Phi = \pm 1$ in that plane (FIG.~\ref{figPlaneSample} demonstrates different cross-section for several times). We study patch's statistics of one of states $\pm 1$ based on the final dominant $\Phi$, in other word, we measure patches with the sign of final dominant state of $\Phi$. FIG.~\ref{figPatchStat} depicts the patch's proportional size distribution in time evolution. Please note that in late time steps, all patches are collapsed and consist a one $3d$ big patch, which is also presented in $2d$ cross-sections.
\newcommand{\trimX}{1}
\newcommand{\trimY}{4}
\newcommand{\totScale}{.4}
\begin{figure}
	\centering
	a) \includegraphics[trim=\trimX cm \trimY cm \trimX cm \trimY cm, clip, width=\totScale \linewidth]{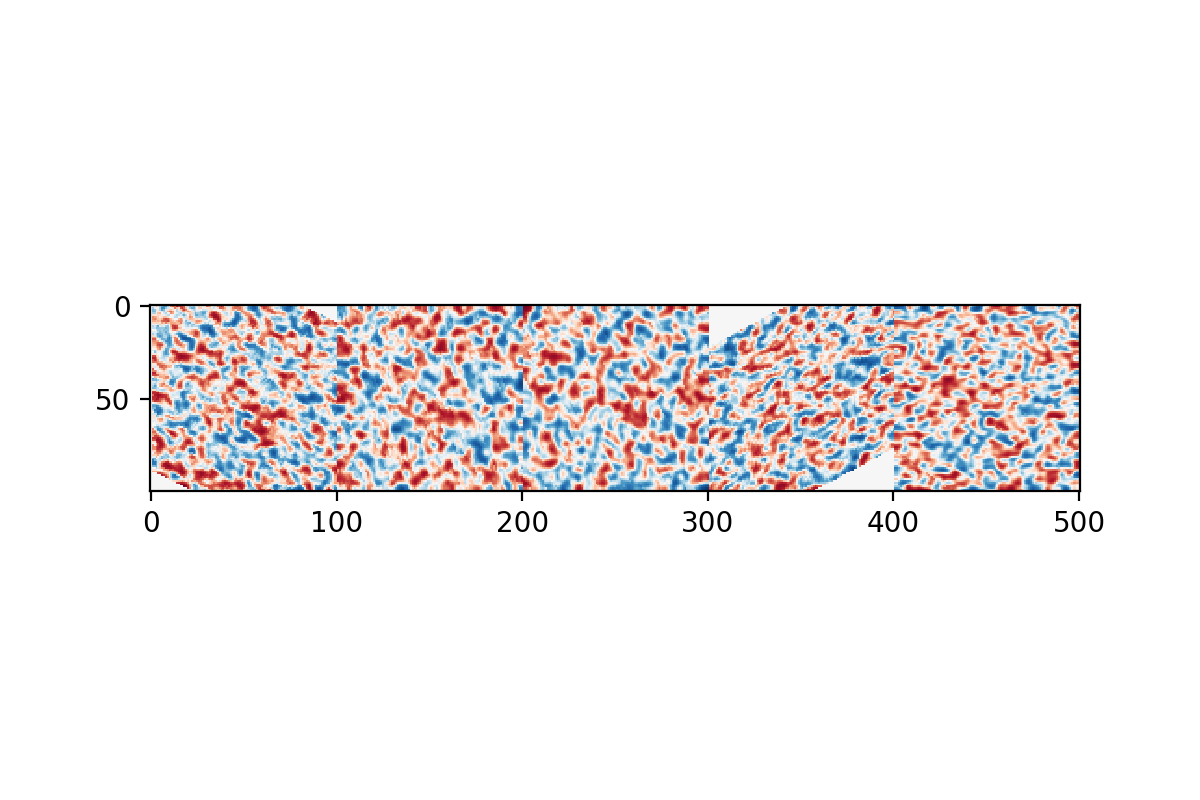}
	b) \includegraphics[trim=\trimX cm \trimY cm \trimX cm \trimY cm, clip, width=\totScale \linewidth]{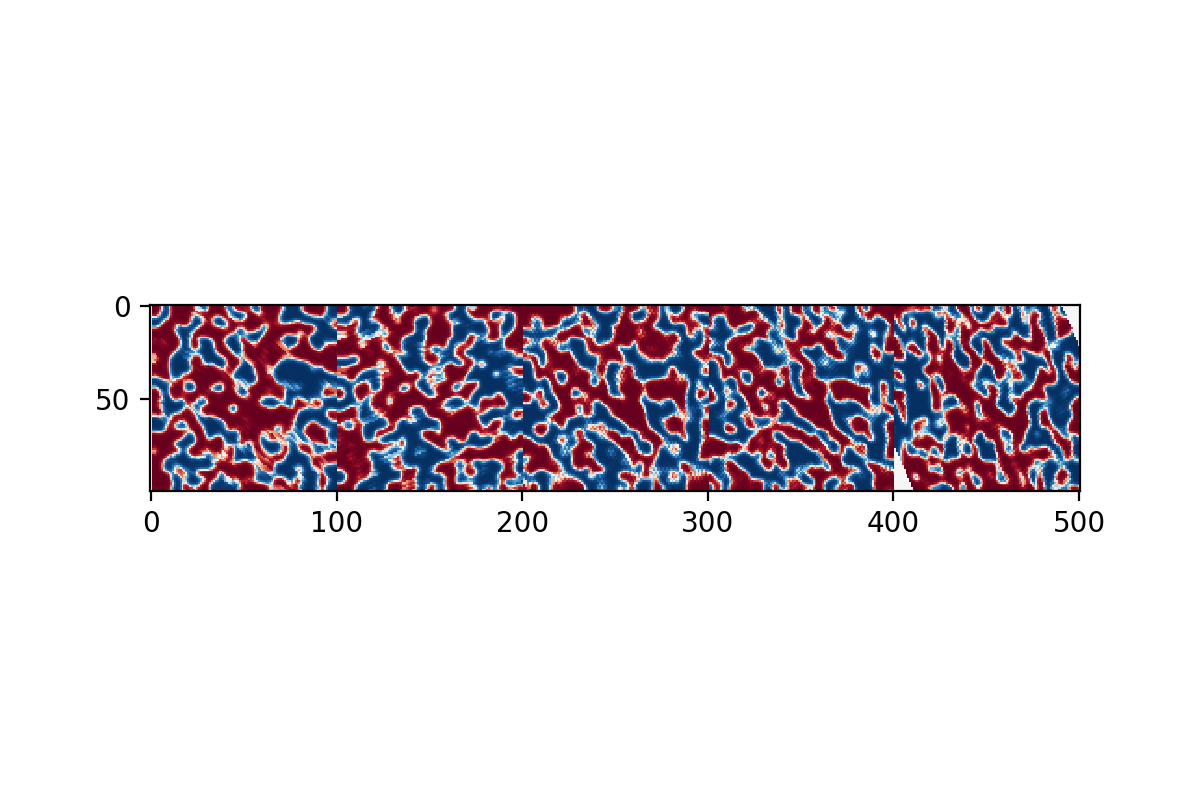}\\
	c) \includegraphics[trim=\trimX cm \trimY cm \trimX cm \trimY cm, clip, width=\totScale \linewidth]{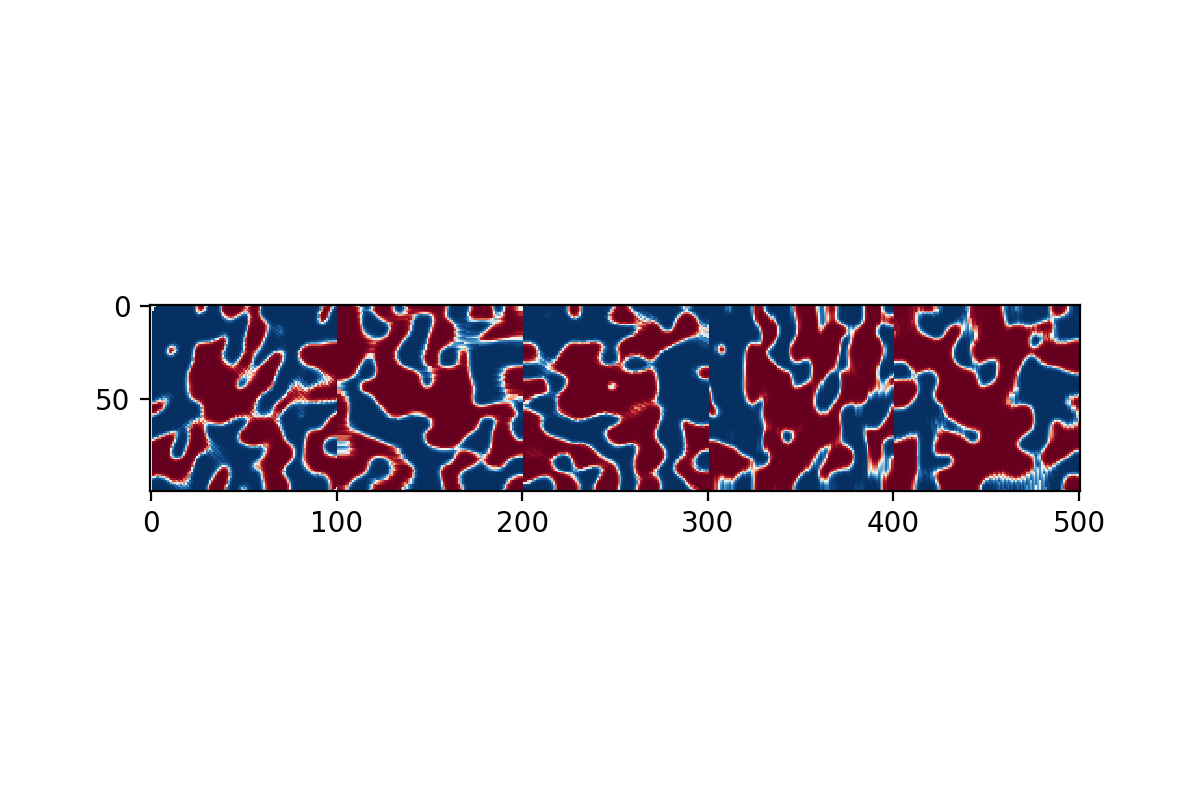}
	d) \includegraphics[trim=\trimX cm \trimY cm \trimX cm \trimY cm, clip, width=\totScale \linewidth]{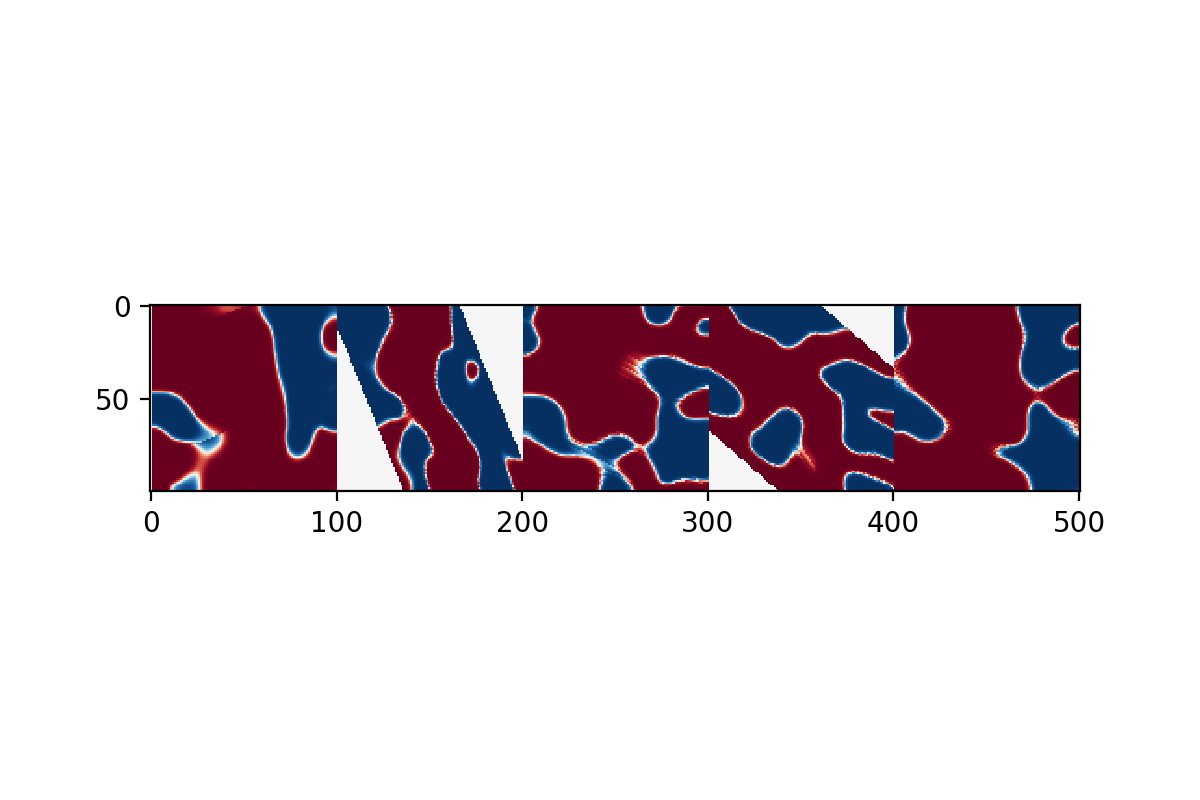} \\
	e) \includegraphics[trim=\trimX cm \trimY cm \trimX cm \trimY cm, clip, width=\totScale \linewidth]{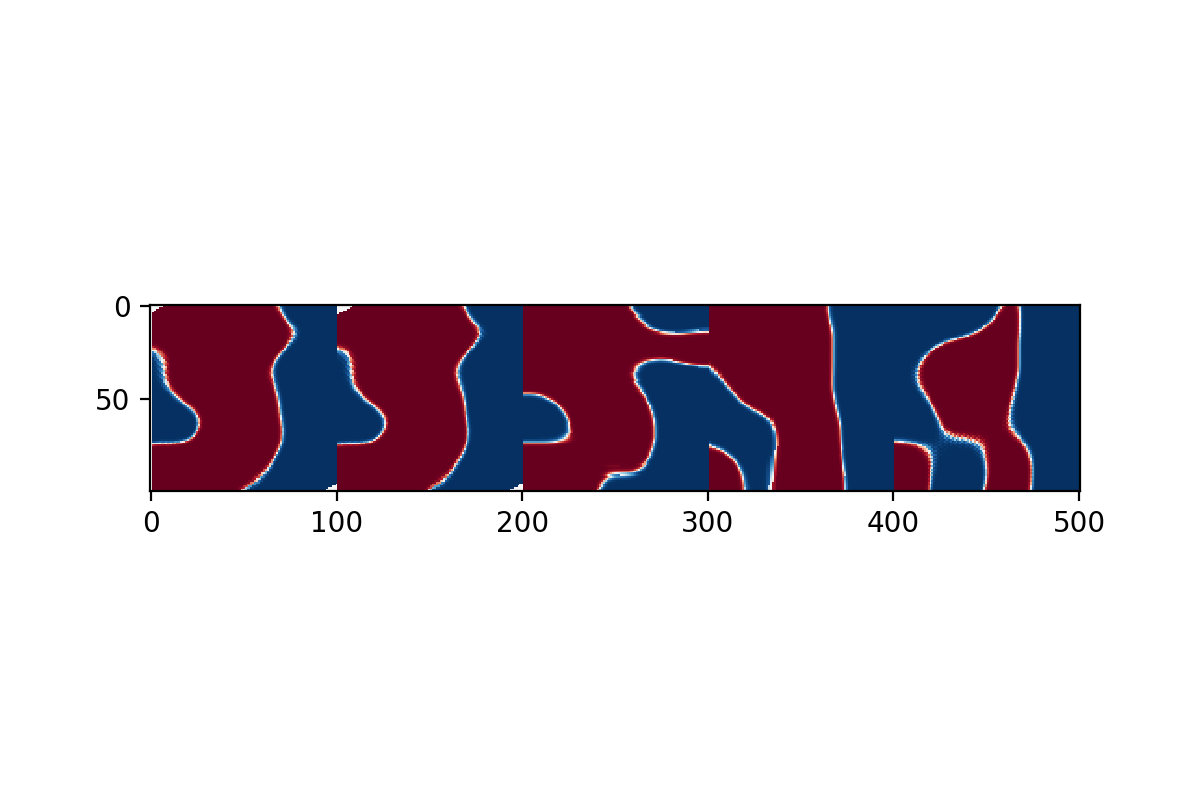}
	f) \includegraphics[trim=\trimX cm \trimY cm \trimX cm \trimY cm, clip, width=\totScale \linewidth]{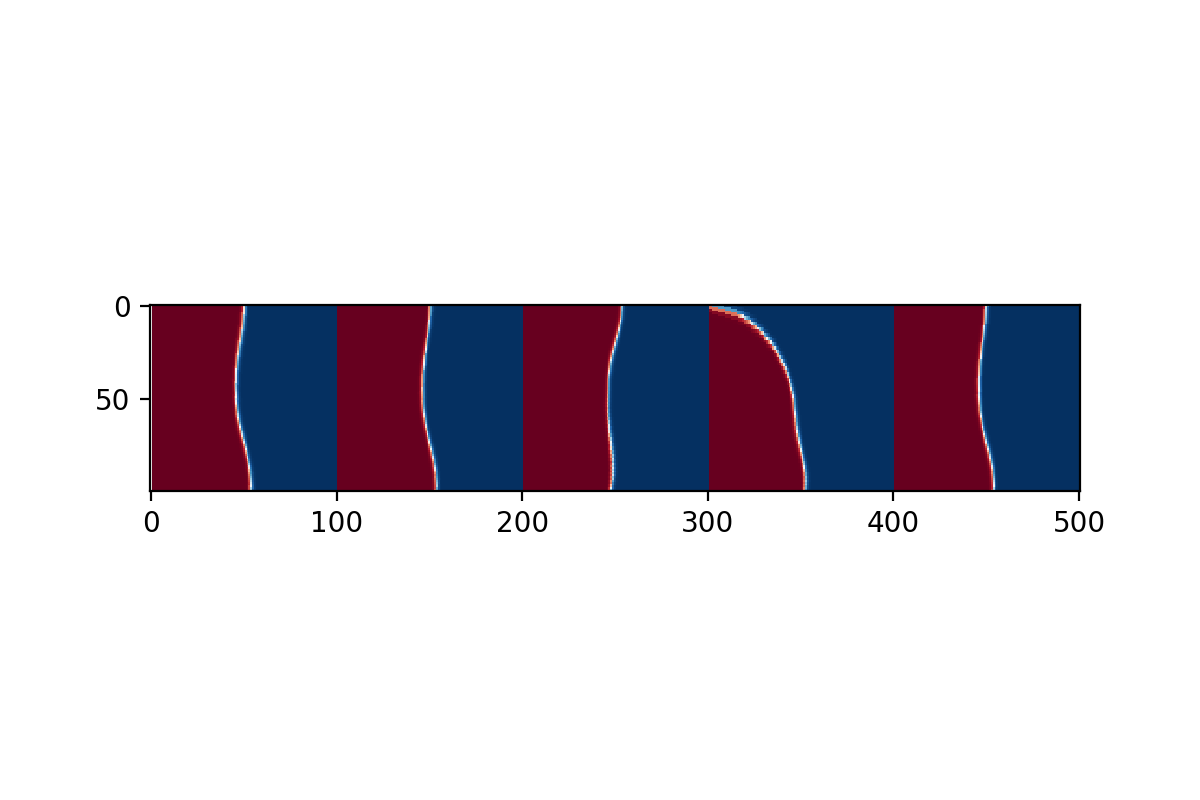} \\
	g) \includegraphics[trim=\trimX cm \trimY cm \trimX cm \trimY cm, clip, width=\totScale \linewidth]{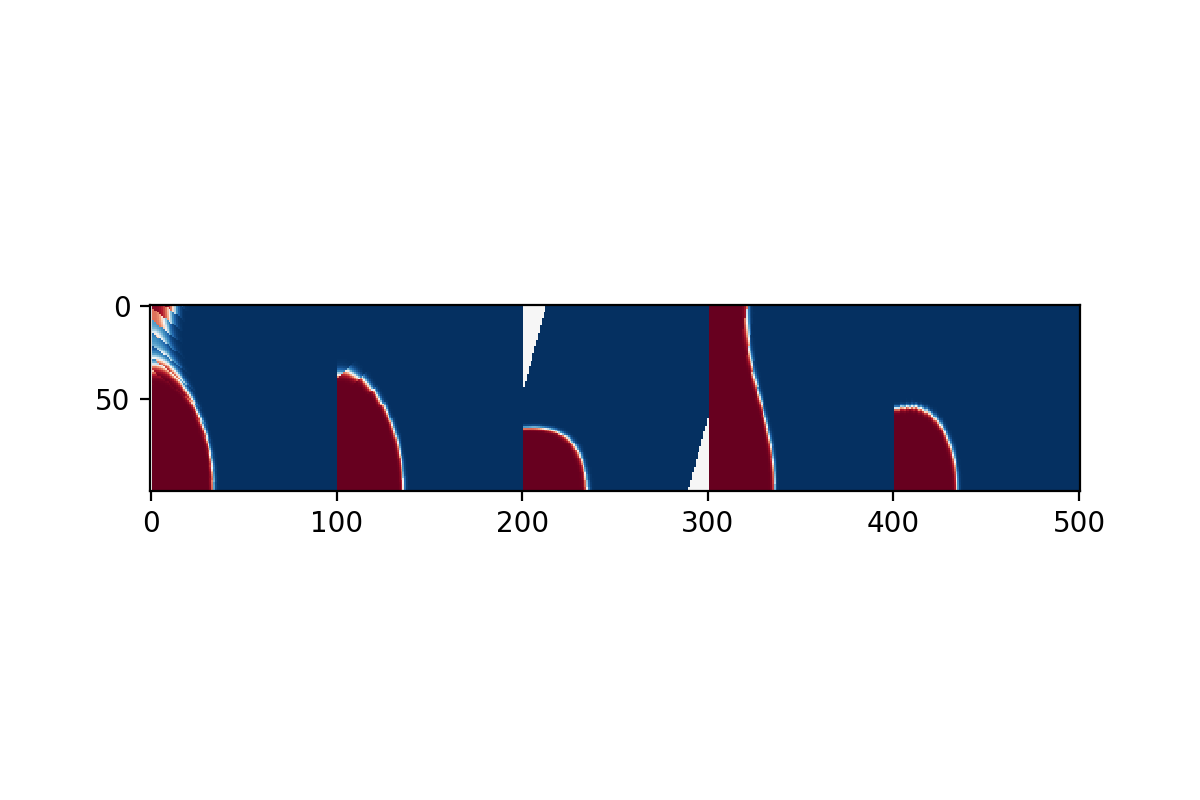}
	h) \includegraphics[trim=\trimX cm \trimY cm \trimX cm \trimY cm, clip, width=\totScale \linewidth]{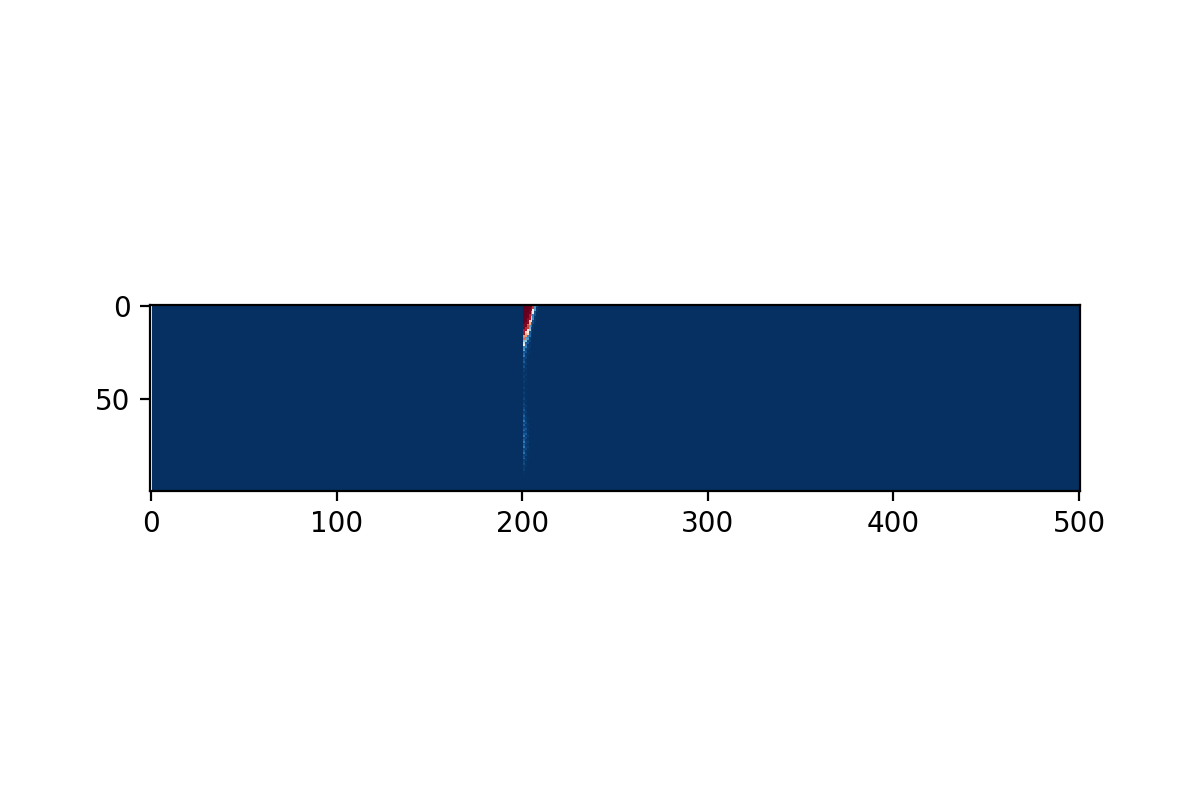}
	
	\caption{$5$ random $2d$ cross-sections from $3d$ cubic lattice in simulation (\ref{equTDGLSim}) for simulation steps a) $15$, b) $39$, c) $158$, d) $630$, e) $1584$, f) $9999$, g) $25118$, h) $39810$. Obviously for early times we have stochastic pattern in the system. As time goes on the patterns start to be formed and finally one state becomes dominant. However before the final state it is obvious to see the appearance of a long wavelength mode in (f).}
	\label{figPlaneSample}
\end{figure}
\begin{figure*}
	\centering
	a) \includegraphics[width=.45\linewidth]{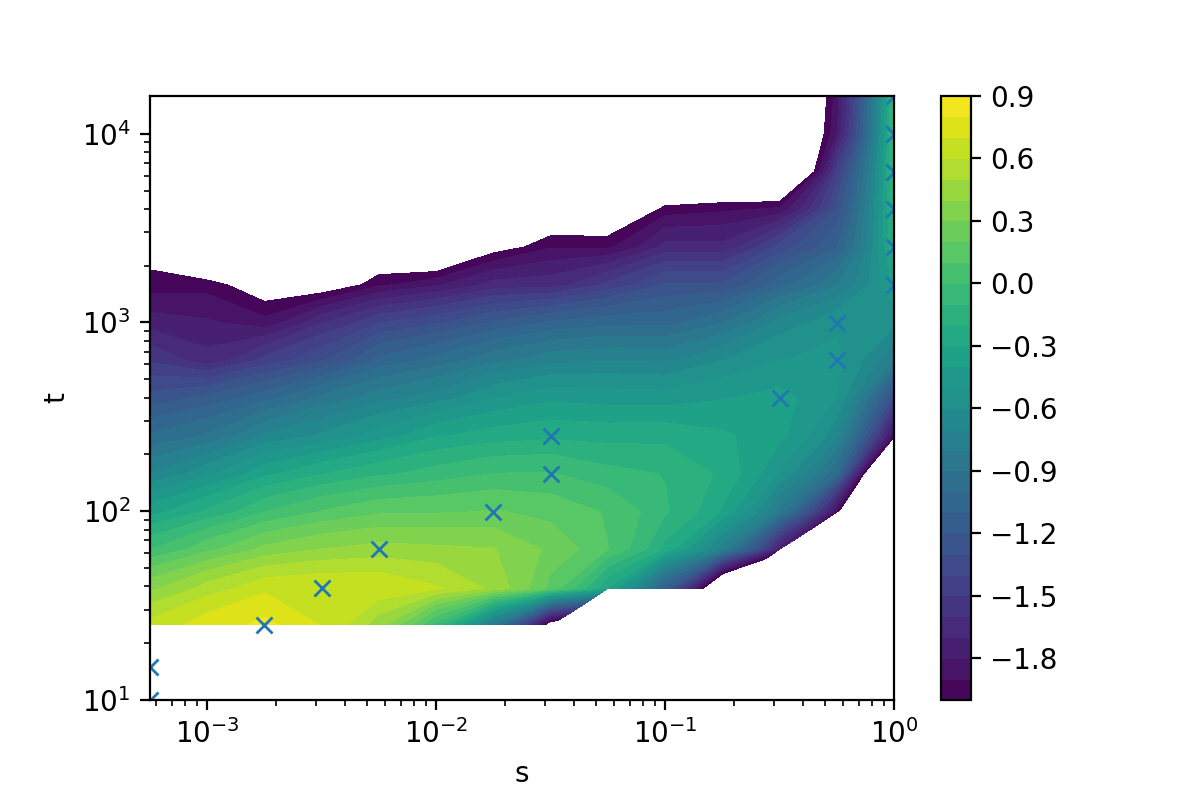}
	b) \includegraphics[width=.45\linewidth]{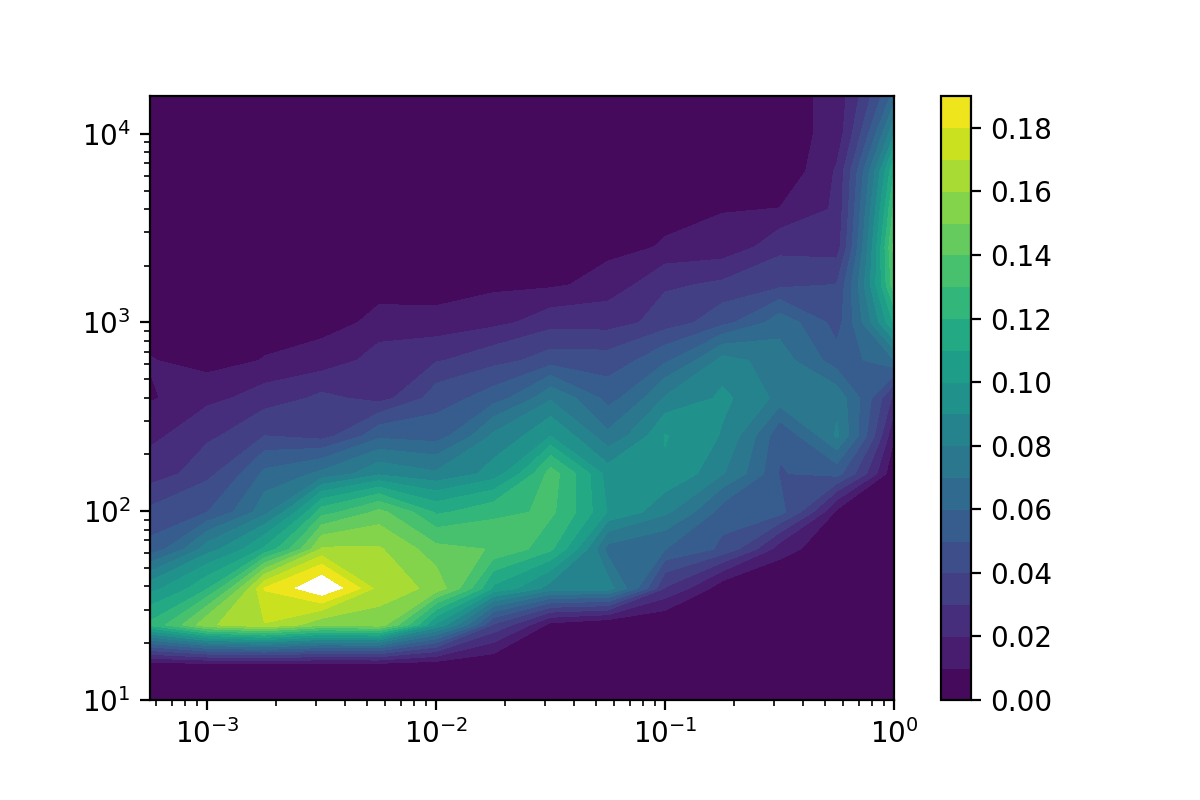} 
	
	\caption{a) Contour plot of patch's proportional size distribution, in several time steps. Color bar also has log scale for better demonstration. Marked points demonstrate the most probable patch's proportional size, in each time step. b) Shows the standard error of patch's proportional size distribution, in an ensemble of simulations with size $30$.}
	\label{figPatchStat}
\end{figure*}
\\

\newcommand{\trimGifX}{4}
\newcommand{\trimGifY}{2.5}
\newcommand{\trimGifWidth}{.45}

\begin{figure}
	\centering
	a) \includegraphics[trim=\trimGifX cm \trimGifY cm \trimGifX cm \trimGifY cm, clip, width=\trimGifWidth \linewidth]{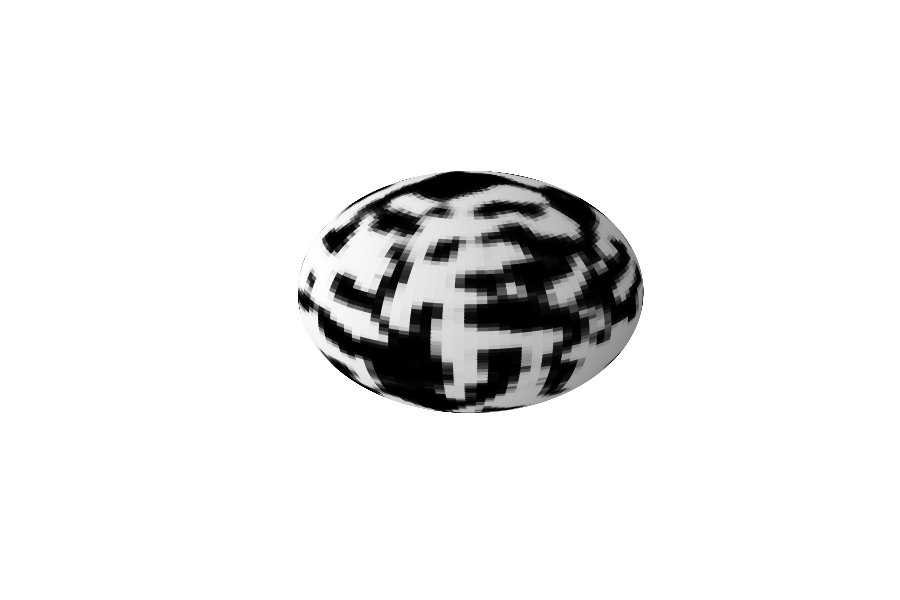}
	b) \includegraphics[trim=\trimGifX cm \trimGifY cm \trimGifX cm \trimGifY cm, clip, width=\trimGifWidth \linewidth]{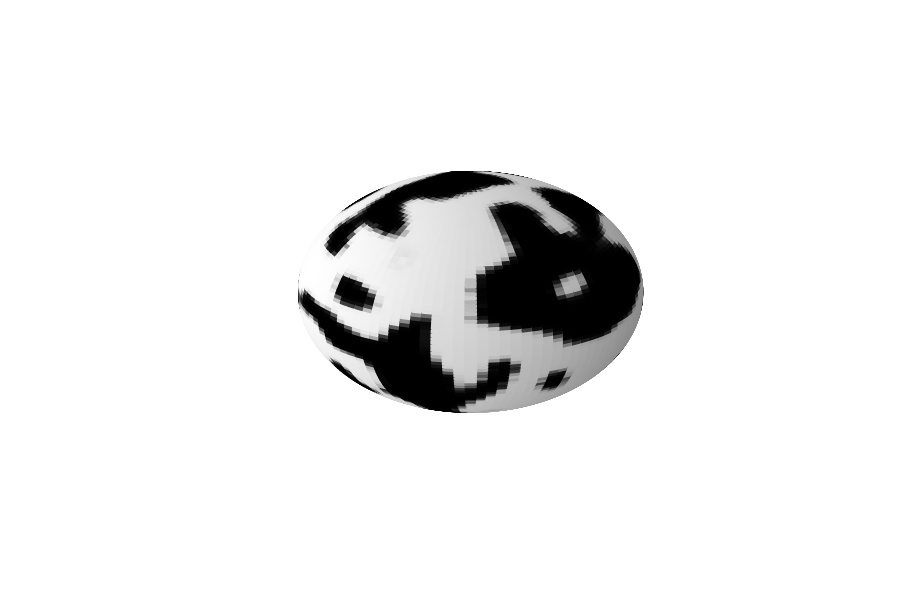} \\
	c) \includegraphics[trim=\trimGifX cm \trimGifY cm \trimGifX cm \trimGifY cm, clip, width=\trimGifWidth \linewidth]{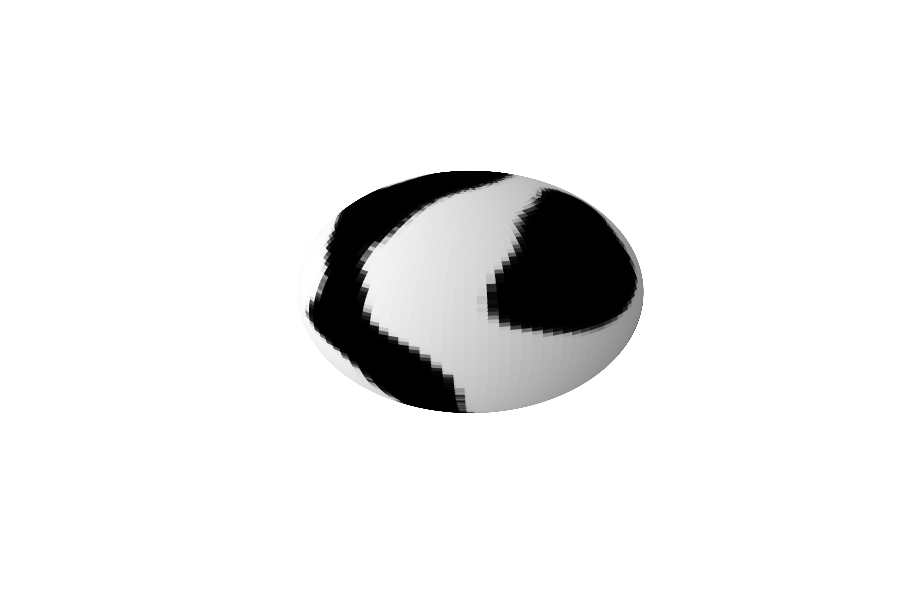}
	d) \includegraphics[trim=\trimGifX cm \trimGifY cm \trimGifX cm \trimGifY cm, clip, width=\trimGifWidth \linewidth]{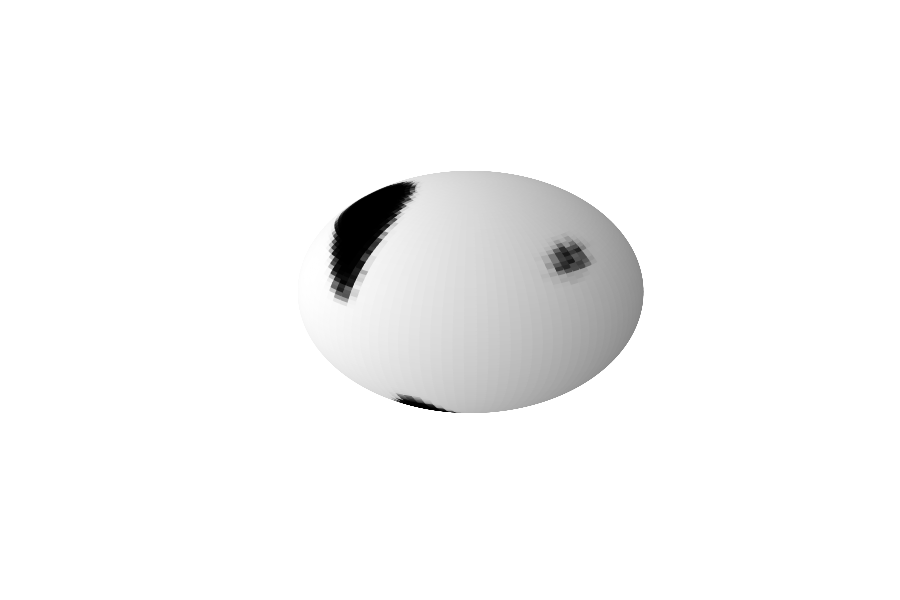}
	\caption{The spherical slices of our TDGL simulation which shows what we have expected. In (a) we see stochastic behavior of patches while with increasing time, the patches became more structured as it is obvious from (b). In later time in (c), the size of patches becomes comparable to the lattice size. And finally we will be close to the final state of TDGL simulation where one of the states becomes dominant. Interestingly in (d), we can see existence of few patches with different sizes. The biggest one can make a dipole and the smaller ones make a framework to think about higher order multi-poles.}
	\label{fig:patch}
\end{figure}

\section{Data sets and more results}\label{data-sets-app}
In this appendix we report which data sets we have used to find the likelihoods for GLTofDE parameters. We have used background data points given in TABLE \ref{back-data}. In addition to background data points we have used CMB distance data point since it is a model independent information about the CMB. We have used $\Omega_m h^2$ as a prior based on Planck 2015 but we see our final results are not sensitive to this prior, see TABLE \ref{table:best-fit} for the results. Since our model is a DE model and not a modified gravity one then we expect that the evolutionary equation for the linear structure i.e. $f\sigma_8(z)$ be affected just by the modification in the background behavior. This means we can also use $f\sigma_8(z)$ data sets from TABLE \ref{tab:data} to constrain our parameters. We have reported the best fits in TABLE \ref{table:best-fit} and likelihoods in FIGS. \ref{fig:contour-back} and \ref{fig:contour-sigma}. In addition we have plotted $H(z)$ versus redshift in FIG. \ref{fig:Hsz} for different values of $\alpha$ to show the general behavior of our model is not sensitive to this parameter too much. However we have plotted $H(z)$ and $f\sigma_8(z)$ versus redshift in FIGS. \ref{fig:Hz} and \ref{fig:fs8} for $\alpha=5$ respectively.

\subsection{A comment on $\Omega^{like}_k$}
As it is obvious from our results in TABLE \ref{table:best-fit} we do get a negative $\Omega^{like}_k$ which seems very far from what we have expected. First we would like to emphasize that since this term, $\Omega^{like}_k$, just appears in the dynamics (i.e. Friedmann equation) and plays no role in the geometrical kinematics (e.g. distances) then it should not be confused with the spatial curvature term. In comparison to \cite{Wang:2018fng}, the difference is in our parametrization. In \cite{Wang:2018fng} the Friedmann equation has been written as 
\begin{eqnarray} \label{omegak}
H^2(z)=H_0^2\big[\Omega_r\,(1+z)^4+\Omega_m\,(1+z)^3+\Omega_{\Lambda}\,Y(z)\big]
\end{eqnarray}
and from the data $Y(z)$ became reconstructed. It means they priory assumed flatness but the price they had to pay is getting negative energy density for dark energy as it is obvious in FIG.1 in \cite{Wang:2018fng}. However we think our parametrization is more physical since we get an always positive energy density for dark energy while we our model predicts exactly the same behavior as their $Y(z)$. We can rebuild their dark energy density $Y(z)$ as\footnote{Note that in \cite{Wang:2018fng} they have $X(z)$ insted of $Y(z)$ but we have used $X(z)$ for a different quantity.} 
\begin{eqnarray} \label{XY}
Y(z)=\big[\Omega^{like}_k(1+z)^2+\Omega_\Lambda\,X(z)\big]/[1-\Omega_m-\Omega_r].
\end{eqnarray}

\begin{figure}[h]
	\centering

	\begingroup
	\makeatletter
	\providecommand\color[2][]{%
		\GenericError{(gnuplot) \space\space\space\@spaces}{%
			Package color not loaded in conjunction with
			terminal option `colourtext'%
		}{See the gnuplot documentation for explanation.%
		}{Either use 'blacktext' in gnuplot or load the package
			color.sty in LaTeX.}%
		\renewcommand\color[2][]{}%
	}%
	\providecommand\includegraphics[2][]{%
		\GenericError{(gnuplot) \space\space\space\@spaces}{%
			Package graphicx or graphics not loaded%
		}{See the gnuplot documentation for explanation.%
		}{The gnuplot epslatex terminal needs graphicx.sty or graphics.sty.}%
		\renewcommand\includegraphics[2][]{}%
	}%
	\providecommand\rotatebox[2]{#2}%
	\@ifundefined{ifGPcolor}{%
		\newif\ifGPcolor
		\GPcolortrue
	}{}%
	\@ifundefined{ifGPblacktext}{%
		\newif\ifGPblacktext
		\GPblacktexttrue
	}{}%
	\let\gplgaddtomacro\g@addto@macro
	\gdef\gplbacktext{}%
	\gdef\gplfronttext{}%
	\makeatother
	\ifGPblacktext
	\def\colorrgb#1{}%
	\def\colorgray#1{}%
	\else
	\ifGPcolor
	\def\colorrgb#1{\color[rgb]{#1}}%
	\def\colorgray#1{\color[gray]{#1}}%
	\expandafter\def\csname LTw\endcsname{\color{white}}%
	\expandafter\def\csname LTb\endcsname{\color{black}}%
	\expandafter\def\csname LTa\endcsname{\color{black}}%
	\expandafter\def\csname LT0\endcsname{\color[rgb]{1,0,0}}%
	\expandafter\def\csname LT1\endcsname{\color[rgb]{0,1,0}}%
	\expandafter\def\csname LT2\endcsname{\color[rgb]{0,0,1}}%
	\expandafter\def\csname LT3\endcsname{\color[rgb]{1,0,1}}%
	\expandafter\def\csname LT4\endcsname{\color[rgb]{0,1,1}}%
	\expandafter\def\csname LT5\endcsname{\color[rgb]{1,1,0}}%
	\expandafter\def\csname LT6\endcsname{\color[rgb]{0,0,0}}%
	\expandafter\def\csname LT7\endcsname{\color[rgb]{1,0.3,0}}%
	\expandafter\def\csname LT8\endcsname{\color[rgb]{0.5,0.5,0.5}}%
	\else
	\def\colorrgb#1{\color{black}}%
	\def\colorgray#1{\color[gray]{#1}}%
	\expandafter\def\csname LTw\endcsname{\color{white}}%
	\expandafter\def\csname LTb\endcsname{\color{black}}%
	\expandafter\def\csname LTa\endcsname{\color{black}}%
	\expandafter\def\csname LT0\endcsname{\color{black}}%
	\expandafter\def\csname LT1\endcsname{\color{black}}%
	\expandafter\def\csname LT2\endcsname{\color{black}}%
	\expandafter\def\csname LT3\endcsname{\color{black}}%
	\expandafter\def\csname LT4\endcsname{\color{black}}%
	\expandafter\def\csname LT5\endcsname{\color{black}}%
	\expandafter\def\csname LT6\endcsname{\color{black}}%
	\expandafter\def\csname LT7\endcsname{\color{black}}%
	\expandafter\def\csname LT8\endcsname{\color{black}}%
	\fi
	\fi
	\setlength{\unitlength}{0.0500bp}%
	\ifx\gptboxheight\undefined%
	\newlength{\gptboxheight}%
	\newlength{\gptboxwidth}%
	\newsavebox{\gptboxtext}%
	\fi%
	\setlength{\fboxrule}{0.5pt}%
	\setlength{\fboxsep}{1pt}%
	\begin{picture}(5100.00,3400.00)%
	\gplgaddtomacro\gplbacktext{%
		\csname LTb\endcsname
		\put(747,595){\makebox(0,0)[r]{\strut{}$-1$}}%
		\csname LTb\endcsname
		\put(747,1031){\makebox(0,0)[r]{\strut{}$-0.5$}}%
		\csname LTb\endcsname
		\put(747,1468){\makebox(0,0)[r]{\strut{}$0$}}%
		\csname LTb\endcsname
		\put(747,1904){\makebox(0,0)[r]{\strut{}$0.5$}}%
		\csname LTb\endcsname
		\put(747,2340){\makebox(0,0)[r]{\strut{}$1$}}%
		\csname LTb\endcsname
		\put(747,2777){\makebox(0,0)[r]{\strut{}$1.5$}}%
		\csname LTb\endcsname
		\put(747,3213){\makebox(0,0)[r]{\strut{}$2$}}%
		\csname LTb\endcsname
		\put(849,409){\makebox(0,0){\strut{}$0$}}%
		\csname LTb\endcsname
		\put(1506,409){\makebox(0,0){\strut{}$0.5$}}%
		\csname LTb\endcsname
		\put(2164,409){\makebox(0,0){\strut{}$1$}}%
		\csname LTb\endcsname
		\put(2821,409){\makebox(0,0){\strut{}$1.5$}}%
		\csname LTb\endcsname
		\put(3478,409){\makebox(0,0){\strut{}$2$}}%
		\csname LTb\endcsname
		\put(4136,409){\makebox(0,0){\strut{}$2.5$}}%
		\csname LTb\endcsname
		\put(4793,409){\makebox(0,0){\strut{}$3$}}%
	}%
	\gplgaddtomacro\gplfronttext{%
		\csname LTb\endcsname
		\put(153,1904){\rotatebox{-270}{\makebox(0,0){\strut{}$Y(z)$}}}%
		\csname LTb\endcsname
		\put(2821,130){\makebox(0,0){\strut{}$z$}}%
		\csname LTb\endcsname
		\put(4005,3046){\makebox(0,0)[r]{\strut{}Background and $f\sigma_8$}}%
		\csname LTb\endcsname
		\put(4005,2860){\makebox(0,0)[r]{\strut{}Background}}%
	}%
	\gplbacktext
	\put(0,0){\includegraphics{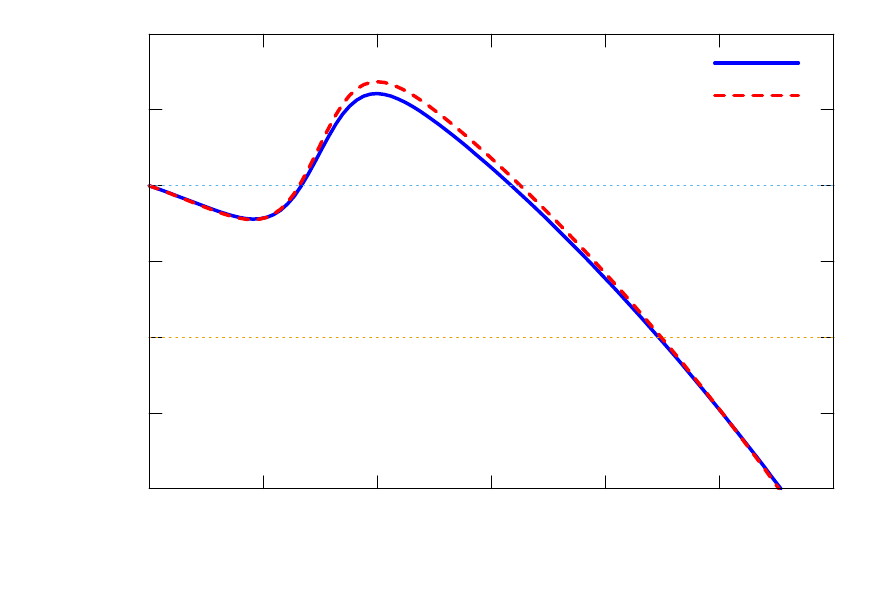}}%
	\gplfronttext
	\end{picture}%
	\endgroup

	\caption{In this figure we have plotted $Y(z)$ which is defined in relation (\ref{XY}) versus redshift for both best fits of our parameters in TABLE \ref{table:best-fit}. This function is what has been named $X(z)$ and be reconstructed from data in \cite{Wang:2018fng}. By comparing our prediction with FIG.1 in \cite{Wang:2018fng} it is obvious that our model behaves as data wants. We do get a negative $Y(z)$ for redshifts above $z\sim 2.2$ and a maximum around $z\sim 1$ exactly same as \cite{Wang:2018fng}.}
	\label{fig:XY}
\end{figure}

In FIG. \ref{fig:XY} 
we have plotted the above $Y(z)$ with our best fits which exactly shows a same behavior as FIG.1 in \cite{Wang:2018fng}. This means GLTofDE could predict the data even with less data set. this negative cosmological constant is also reported in \cite{Dutta:2018vmq}. We think our re-parametrization may give a clue to look for departure from spatially flatness geometry which remains for the future works. If it is true then we do not need to assume negative energy density for the dark energy which does not seem physically interesting.

\begin{table*}[t]
	\begin{tabular}{ |c|c|c|c|c|c|c|c|c| }
		\hline
		CMB & BAO & BAO  & BAO\\ \hline
		&  & &   \\
		CMB first peak \cite{Aghanim:2018eyx} & BOSS DR12 ($z=0.38$) \cite{Alam:2016hwk}  & BOSS DR12 ($z=0.51$) \cite{Alam:2016hwk}  &BOSS DR12 ($z=0.61$) \cite{Alam:2016hwk} \\
		&  & &   \\
		$100 \Theta = 1.04085 \pm 0.00047$	& $H(z)/(1+z)=59.05\pm 1.38$ & $H(z)/(1+z) = 59.87\pm 1.26$ & $H(z)/(1+z)=60.43 \pm 1.3$ \\
		&&&\\
		\hline
		Hubble & Quasars & BAO  & BAO   \\
		\hline
		&  & &    \\
		Local $H_0$ \cite{riess18} & BOSS DR14 ($z=1.52$) \cite{Zarrouk:2018vwy}  &BOSS Ly-$\alpha$ ($z=2.33$) \cite{Bautista:2017zgn}  & BOSS Ly-$\alpha$ ($z=2.40$) \cite{Bourboux:2017cbm} \\
		&  & &   \\
		$H_0=73.48\pm1.66$ & $H(z)/(1+z)= 63.1\pm 4.96$ & $H(z)/(1+z)= 67.27\pm 2.40$  &$H(z)/(1+z) = 67.14 \pm 1.65$  \\
		&  & &  \\
		\hline
		BAO & BAO & BAO & BAO\\ \hline
		&  & &\\
		6dFGS ($z=0.106$)\cite{6df} & DES ($z=0.81$)\cite{Abbott:2017wcz} & MGS ($z=0.150$)\cite{Ross:2014qpa} & WiggleZ ($z=0.44$)\cite{Kazin:2014qga} \\
		& & &\\
		$D_V=449.1\pm20.1$ & $D_M=2861.4\pm115.2$ & $D_V=657.7\pm25.6$ & $D_V=1698.7\pm82.0$\\
		& & &\\
		\hline 
	    BAO	& BAO & BAO & BAO \\
		\hline 
		&  & & \\
		WiggleZ ($z=0.60$)\cite{Kazin:2014qga} & WiggleZ ($z=0.73$)\cite{Kazin:2014qga} & DR14 LRG ($z=0.72$)\cite{Bautista:2017wwp} & BOSS Ly-$\alpha$ ($z=2.40$)\cite{Bourboux:2017cbm}\\
		& & &\\
		$D_V=2200.0\pm100.3$ & $D_V=2491.0\pm85.6$ & $D_V=2340.4\pm62.6$ & $D_M=5378.1\pm179.3$ \\
		&&& \\
 		\hline
		
	\end{tabular}
	\caption{\label{tab:data}The background dataset. $\Theta$ represents the distance of the last scattering surface to us. We also use thirteen BAO measurements and one data point from BOSS DR14 quasars. The additional data point is the Hubble parameter at the present time, $H_0$, which  is reported by analysis of supernovae. All values of $H$ and distances in the above table are in units of km/s/Mpc and Mpc respectively. In addition, we do our $\chi^2$ analysis with a prior on $\Omega_m h^2$ given by Planck 2018.}\label{back-data}
\end{table*}

\begin{table*}[t]
	\begin{tabular}{ |c|c|c|c|c|c|c|c|c| }
		\hline
		6dFGS+SnIa \cite{Huterer:2016uyq} & SDSS-MGS \cite{Howlett:2014opa} & SDSS-LRG \cite{Samushia:2011cs}  \\ \hline
		&  &   \\
		$0.428\pm 0.0465\,\,(z=0.02)$ & $0.490\pm 0.145\,\,(z=0.15)$ & $0.3512\pm 0.0583\,\,(z=0.25)$   \\ 
		&  &   \\ \hline
		BOSS-LOWZ \cite{Sanchez:2013tga} &SDSS-CMASS \cite{Chuang:2013wga}  & WiggleZ \cite{Blake:2012pj}    \\
		\hline
		&  &     \\
		$0.384\pm 0.095\,\,(z=0.32)$ & $0.488\pm 0.060\,\,(z=0.59)$ &$0.413\pm 0.080\,\,(z=0.44)$ \\
		&  &   \\ 
		\hline
		WiggleZ \cite{Blake:2012pj} &WiggleZ \cite{Blake:2012pj}  & Vipers PDR-2 \cite{Pezzotta:2016gbo}    \\ \hline
		&  &     \\
		$0.390\pm 0.063\,\,(z=0.60)$ & $0.437\pm 0.072\,\,(z=0.73)$ &$0.400\pm 0.110\,\,(z=0.86)$ \\
		&  &     \\
		\hline
	\end{tabular}
	\caption{\label{tab:data}$f\sigma_8$ Datasets.}\label{fsigma-data}
\end{table*}

\begin{figure}
	\centering
	\includegraphics[width=.9\linewidth]{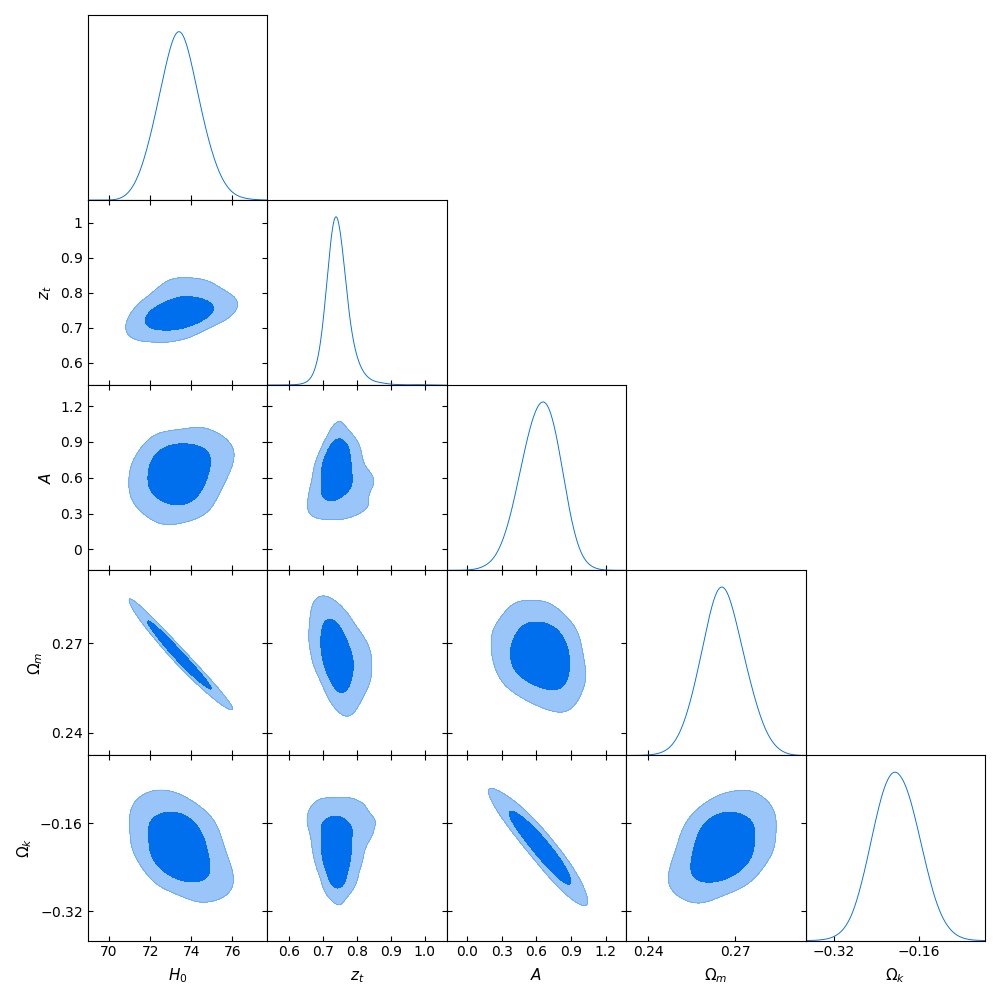}
	\caption{Likelihood of GLTofDE free parameters for $\alpha=5$ if we use just background data points in TABLE \ref{back-data}.}
	\label{fig:contour-back}
\end{figure}

\begin{figure}
	\centering
	\includegraphics[width=.9\linewidth]{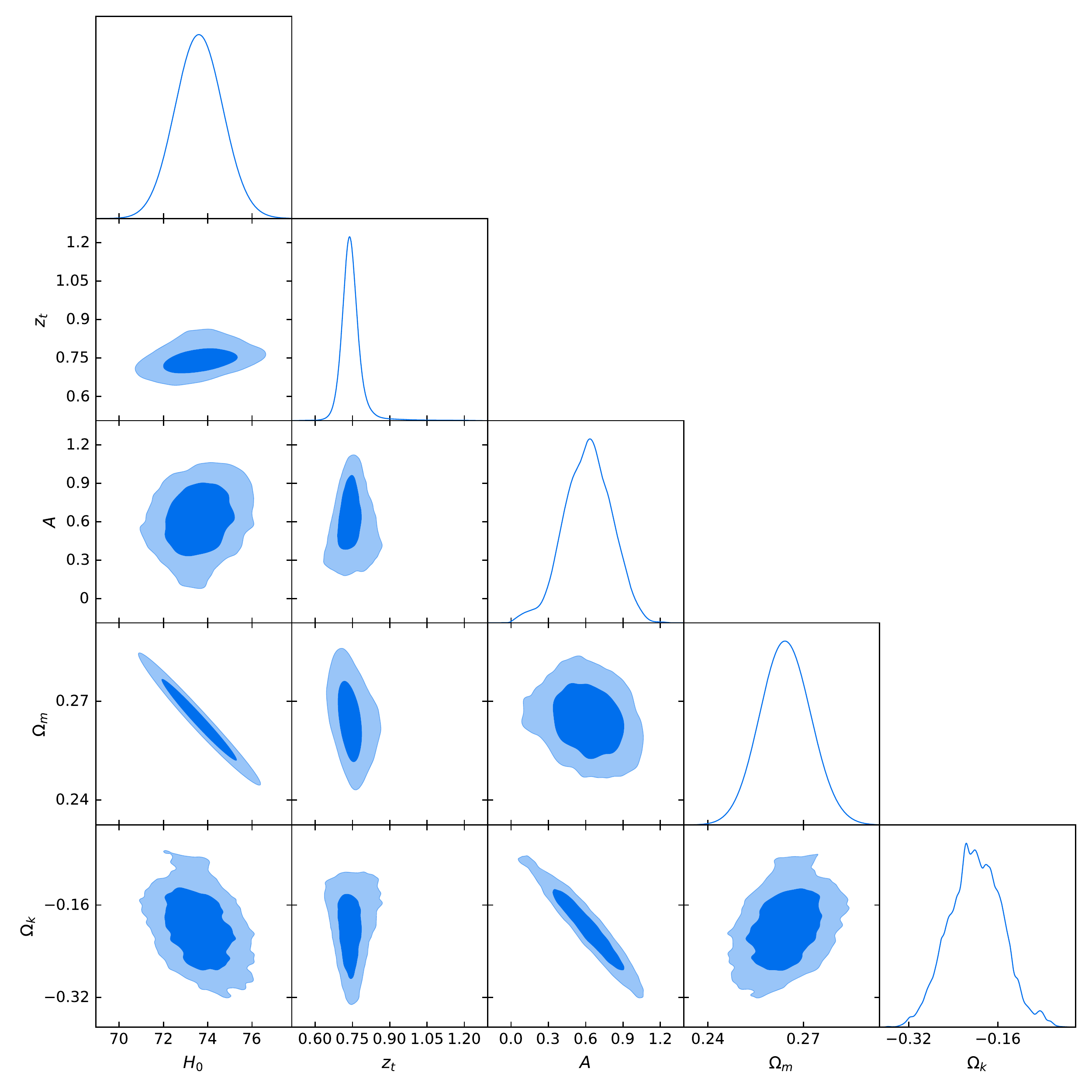}
	\caption{Likelihood of GLTofDE free parameters for $\alpha=5$ if we use both background and $f\sigma_8(z)$ data sets in TABLES. \ref{back-data} and \ref{fsigma-data}.}
	\label{fig:contour-sigma}
\end{figure}

\begin{figure}
	\centering
	\includegraphics[width=.9\linewidth]{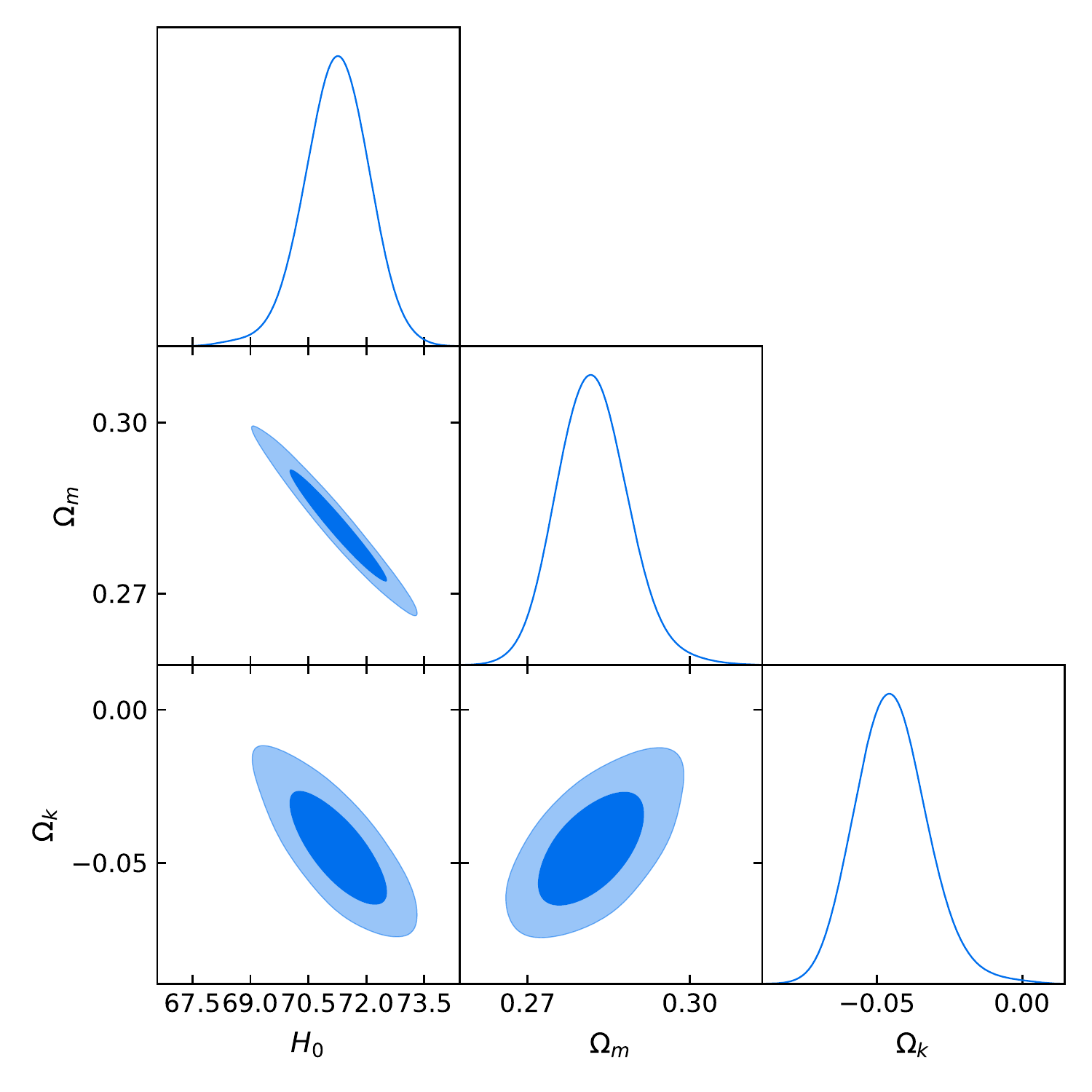}
	\caption{Likelihood of $\Lambda$CDM free parameters if we use both background and $f\sigma_8(z)$ data sets in TABLES. \ref{back-data} and \ref{fsigma-data}.}
	\label{fig:contour-sigma}
\end{figure}

\begin{figure}
	\centering
	\includegraphics[width=.9\linewidth]{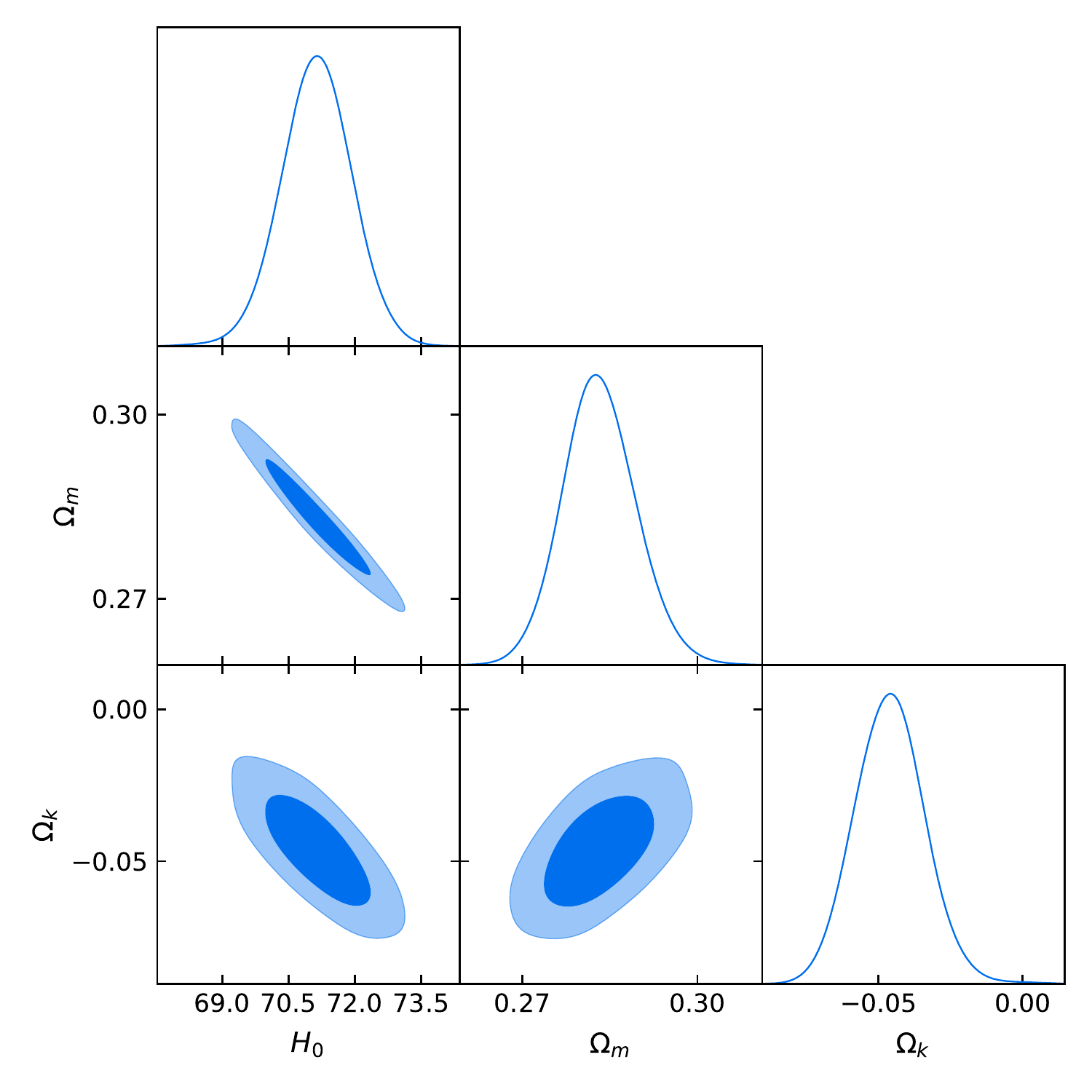}
	\caption{Likelihood of $\Lambda$CDM free parameters if we use only background data set in TABLE \ref{back-data}.}
	\label{fig:contour-sigma}
\end{figure}

\begin{figure}[h]
	\centering
	\begingroup
	\makeatletter
	\providecommand\color[2][]{%
		\GenericError{(gnuplot) \space\space\space\@spaces}{%
			Package color not loaded in conjunction with
			terminal option `colourtext'%
		}{See the gnuplot documentation for explanation.%
		}{Either use 'blacktext' in gnuplot or load the package
			color.sty in LaTeX.}%
		\renewcommand\color[2][]{}%
	}%
	\providecommand\includegraphics[2][]{%
		\GenericError{(gnuplot) \space\space\space\@spaces}{%
			Package graphicx or graphics not loaded%
		}{See the gnuplot documentation for explanation.%
		}{The gnuplot epslatex terminal needs graphicx.sty or graphics.sty.}%
		\renewcommand\includegraphics[2][]{}%
	}%
	\providecommand\rotatebox[2]{#2}%
	\@ifundefined{ifGPcolor}{%
		\newif\ifGPcolor
		\GPcolortrue
	}{}%
	\@ifundefined{ifGPblacktext}{%
		\newif\ifGPblacktext
		\GPblacktexttrue
	}{}%
	\let\gplgaddtomacro\g@addto@macro
	\gdef\gplbacktext{}%
	\gdef\gplfronttext{}%
	\makeatother
	\ifGPblacktext
	\def\colorrgb#1{}%
	\def\colorgray#1{}%
	\else
	\ifGPcolor
	\def\colorrgb#1{\color[rgb]{#1}}%
	\def\colorgray#1{\color[gray]{#1}}%
	\expandafter\def\csname LTw\endcsname{\color{white}}%
	\expandafter\def\csname LTb\endcsname{\color{black}}%
	\expandafter\def\csname LTa\endcsname{\color{black}}%
	\expandafter\def\csname LT0\endcsname{\color[rgb]{1,0,0}}%
	\expandafter\def\csname LT1\endcsname{\color[rgb]{0,1,0}}%
	\expandafter\def\csname LT2\endcsname{\color[rgb]{0,0,1}}%
	\expandafter\def\csname LT3\endcsname{\color[rgb]{1,0,1}}%
	\expandafter\def\csname LT4\endcsname{\color[rgb]{0,1,1}}%
	\expandafter\def\csname LT5\endcsname{\color[rgb]{1,1,0}}%
	\expandafter\def\csname LT6\endcsname{\color[rgb]{0,0,0}}%
	\expandafter\def\csname LT7\endcsname{\color[rgb]{1,0.3,0}}%
	\expandafter\def\csname LT8\endcsname{\color[rgb]{0.5,0.5,0.5}}%
	\else
	\def\colorrgb#1{\color{black}}%
	\def\colorgray#1{\color[gray]{#1}}%
	\expandafter\def\csname LTw\endcsname{\color{white}}%
	\expandafter\def\csname LTb\endcsname{\color{black}}%
	\expandafter\def\csname LTa\endcsname{\color{black}}%
	\expandafter\def\csname LT0\endcsname{\color{black}}%
	\expandafter\def\csname LT1\endcsname{\color{black}}%
	\expandafter\def\csname LT2\endcsname{\color{black}}%
	\expandafter\def\csname LT3\endcsname{\color{black}}%
	\expandafter\def\csname LT4\endcsname{\color{black}}%
	\expandafter\def\csname LT5\endcsname{\color{black}}%
	\expandafter\def\csname LT6\endcsname{\color{black}}%
	\expandafter\def\csname LT7\endcsname{\color{black}}%
	\expandafter\def\csname LT8\endcsname{\color{black}}%
	\fi
	\fi
	\setlength{\unitlength}{0.0500bp}%
	\ifx\gptboxheight\undefined%
	\newlength{\gptboxheight}%
	\newlength{\gptboxwidth}%
	\newsavebox{\gptboxtext}%
	\fi%
	\setlength{\fboxrule}{0.5pt}%
	\setlength{\fboxsep}{1pt}%
	\begin{picture}(5100.00,3400.00)%
	\gplgaddtomacro\gplbacktext{%
		\csname LTb\endcsname
		\put(362,396){\makebox(0,0)[r]{\strut{}$55$}}%
		\csname LTb\endcsname
		\put(362,972){\makebox(0,0)[r]{\strut{}$60$}}%
		\csname LTb\endcsname
		\put(362,1548){\makebox(0,0)[r]{\strut{}$65$}}%
		\csname LTb\endcsname
		\put(362,2123){\makebox(0,0)[r]{\strut{}$70$}}%
		\csname LTb\endcsname
		\put(362,2699){\makebox(0,0)[r]{\strut{}$75$}}%
		\csname LTb\endcsname
		\put(362,3275){\makebox(0,0)[r]{\strut{}$80$}}%
		\csname LTb\endcsname
		\put(430,272){\makebox(0,0){\strut{}$0$}}%
		\csname LTb\endcsname
		\put(1289,272){\makebox(0,0){\strut{}$0.5$}}%
		\csname LTb\endcsname
		\put(2147,272){\makebox(0,0){\strut{}$1$}}%
		\csname LTb\endcsname
		\put(3006,272){\makebox(0,0){\strut{}$1.5$}}%
		\csname LTb\endcsname
		\put(3865,272){\makebox(0,0){\strut{}$2$}}%
		\csname LTb\endcsname
		\put(4723,272){\makebox(0,0){\strut{}$2.5$}}%
	}%
	\gplgaddtomacro\gplfronttext{%
		\csname LTb\endcsname
		\put(0,1835){\rotatebox{-270}{\makebox(0,0){\strut{}$H(z)/(1+z)\rm~{[kms^{-1}Mpc^{-1}]}$}}}%
		\begin{Large}
		\csname LTb\endcsname
		\put(2662,86){\makebox(0,0){\strut{}$z$}}%
		\end{Large}
		\begin{footnotesize}
		\csname LTb\endcsname
		\put(4370,3164){\makebox(0,0)[r]{\strut{}$\alpha=4$}}%
		\csname LTb\endcsname
		\put(4370,3040){\makebox(0,0)[r]{\strut{}$\alpha=5$}}%
		\csname LTb\endcsname
		\put(4370,2916){\makebox(0,0)[r]{\strut{}$\alpha=6$}}%
		\csname LTb\endcsname
		\put(4370,2792){\makebox(0,0)[r]{\strut{}$\alpha=7$}}%
		\csname LTb\endcsname
		\put(4370,2668){\makebox(0,0)[r]{\strut{}$\alpha=8$}}%
		\csname LTb\endcsname
		\put(4370,2544){\makebox(0,0)[r]{\strut{}$\alpha=10$}}%
		\csname LTb\endcsname
		\put(4370,2420){\makebox(0,0)[r]{\strut{}$\alpha=20$}}%
		\csname LTb\endcsname
		\put(4370,2296){\makebox(0,0)[r]{\strut{}$\alpha=50$}}%
		\csname LTb\endcsname
		\put(1300,2488){\makebox(0,0)[r]{\strut{}Riess et al.}}%
		\csname LTb\endcsname
		\put(1800,588){\makebox(0,0)[r]{\strut{}BOSS DR12}}%
		\csname LTb\endcsname
		\put(3600,650){\makebox(0,0)[r]{\strut{}DR14 quasars}}%
		\csname LTb\endcsname
		\put(4750,1400){\makebox(0,0)[r]{\strut{}BOSS Ly-$\alpha$}}%
		\end{footnotesize}
	}%
	\gplbacktext
	\put(0,0){\includegraphics{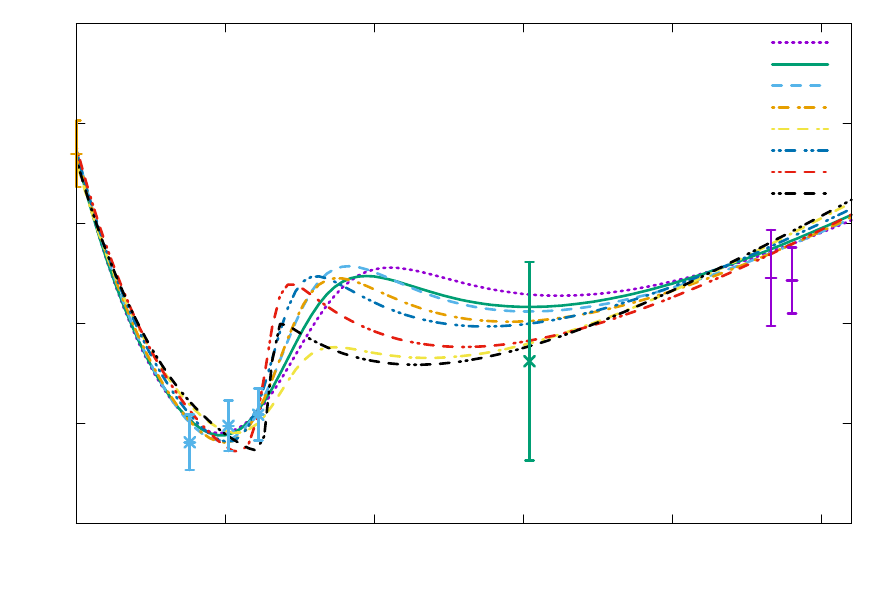}}%
	\gplfronttext
	\end{picture}%
	\endgroup
	
	\caption{We have plotted $H(z)$ in GLTofDE for different values of parameter $\alpha$. $\alpha$ in relation (\ref{X-tanh}) shows the shape of transition and it is obvious from this plot that our model is not very sensitive to its value at least for $H(z)$ data points we have used here. Note that for larger values of $\alpha$ $tanh$-function becomes like a step function.}
	\label{fig:Hsz}
\end{figure}

\begin{figure}[h]
	\centering
		\begingroup
		\makeatletter
		\providecommand\color[2][]{%
			\GenericError{(gnuplot) \space\space\space\@spaces}{%
				Package color not loaded in conjunction with
				terminal option `colourtext'%
			}{See the gnuplot documentation for explanation.%
			}{Either use 'blacktext' in gnuplot or load the package
				color.sty in LaTeX.}%
			\renewcommand\color[2][]{}%
		}%
		\providecommand\includegraphics[2][]{%
			\GenericError{(gnuplot) \space\space\space\@spaces}{%
				Package graphicx or graphics not loaded%
			}{See the gnuplot documentation for explanation.%
			}{The gnuplot epslatex terminal needs graphicx.sty or graphics.sty.}%
			\renewcommand\includegraphics[2][]{}%
		}%
		\providecommand\rotatebox[2]{#2}%
		\@ifundefined{ifGPcolor}{%
			\newif\ifGPcolor
			\GPcolortrue
		}{}%
		\@ifundefined{ifGPblacktext}{%
			\newif\ifGPblacktext
			\GPblacktexttrue
		}{}%
		\let\gplgaddtomacro\g@addto@macro
		\gdef\gplbacktext{}%
		\gdef\gplfronttext{}%
		\makeatother
		\ifGPblacktext
		\def\colorrgb#1{}%
		\def\colorgray#1{}%
		\else
		\ifGPcolor
		\def\colorrgb#1{\color[rgb]{#1}}%
		\def\colorgray#1{\color[gray]{#1}}%
		\expandafter\def\csname LTw\endcsname{\color{white}}%
		\expandafter\def\csname LTb\endcsname{\color{black}}%
		\expandafter\def\csname LTa\endcsname{\color{black}}%
		\expandafter\def\csname LT0\endcsname{\color[rgb]{1,0,0}}%
		\expandafter\def\csname LT1\endcsname{\color[rgb]{0,1,0}}%
		\expandafter\def\csname LT2\endcsname{\color[rgb]{0,0,1}}%
		\expandafter\def\csname LT3\endcsname{\color[rgb]{1,0,1}}%
		\expandafter\def\csname LT4\endcsname{\color[rgb]{0,1,1}}%
		\expandafter\def\csname LT5\endcsname{\color[rgb]{1,1,0}}%
		\expandafter\def\csname LT6\endcsname{\color[rgb]{0,0,0}}%
		\expandafter\def\csname LT7\endcsname{\color[rgb]{1,0.3,0}}%
		\expandafter\def\csname LT8\endcsname{\color[rgb]{0.5,0.5,0.5}}%
		\else
		\def\colorrgb#1{\color{black}}%
		\def\colorgray#1{\color[gray]{#1}}%
		\expandafter\def\csname LTw\endcsname{\color{white}}%
		\expandafter\def\csname LTb\endcsname{\color{black}}%
		\expandafter\def\csname LTa\endcsname{\color{black}}%
		\expandafter\def\csname LT0\endcsname{\color{black}}%
		\expandafter\def\csname LT1\endcsname{\color{black}}%
		\expandafter\def\csname LT2\endcsname{\color{black}}%
		\expandafter\def\csname LT3\endcsname{\color{black}}%
		\expandafter\def\csname LT4\endcsname{\color{black}}%
		\expandafter\def\csname LT5\endcsname{\color{black}}%
		\expandafter\def\csname LT6\endcsname{\color{black}}%
		\expandafter\def\csname LT7\endcsname{\color{black}}%
		\expandafter\def\csname LT8\endcsname{\color{black}}%
		\fi
		\fi
		\setlength{\unitlength}{0.0500bp}%
		\ifx\gptboxheight\undefined%
		\newlength{\gptboxheight}%
		\newlength{\gptboxwidth}%
		\newsavebox{\gptboxtext}%
		\fi%
		\setlength{\fboxrule}{0.5pt}%
		\setlength{\fboxsep}{1pt}%
		\begin{picture}(5100.00,3400.00)%
		\gplgaddtomacro\gplbacktext{%
			\csname LTb\endcsname
			\put(645,595){\makebox(0,0)[r]{\strut{}$0.1$}}%
			\csname LTb\endcsname
			\put(645,969){\makebox(0,0)[r]{\strut{}$0.2$}}%
			\csname LTb\endcsname
			\put(645,1343){\makebox(0,0)[r]{\strut{}$0.3$}}%
			\csname LTb\endcsname
			\put(645,1717){\makebox(0,0)[r]{\strut{}$0.4$}}%
			\csname LTb\endcsname
			\put(645,2091){\makebox(0,0)[r]{\strut{}$0.5$}}%
			\csname LTb\endcsname
			\put(645,2465){\makebox(0,0)[r]{\strut{}$0.6$}}%
			\csname LTb\endcsname
			\put(645,2839){\makebox(0,0)[r]{\strut{}$0.7$}}%
			\csname LTb\endcsname
			\put(645,3213){\makebox(0,0)[r]{\strut{}$0.8$}}%
			\csname LTb\endcsname
			\put(747,409){\makebox(0,0){\strut{}$0$}}%
			\csname LTb\endcsname
			\put(1197,409){\makebox(0,0){\strut{}$0.1$}}%
			\csname LTb\endcsname
			\put(1646,409){\makebox(0,0){\strut{}$0.2$}}%
			\csname LTb\endcsname
			\put(2096,409){\makebox(0,0){\strut{}$0.3$}}%
			\csname LTb\endcsname
			\put(2545,409){\makebox(0,0){\strut{}$0.4$}}%
			\csname LTb\endcsname
			\put(2995,409){\makebox(0,0){\strut{}$0.5$}}%
			\csname LTb\endcsname
			\put(3444,409){\makebox(0,0){\strut{}$0.6$}}%
			\csname LTb\endcsname
			\put(3894,409){\makebox(0,0){\strut{}$0.7$}}%
			\csname LTb\endcsname
			\put(4343,409){\makebox(0,0){\strut{}$0.8$}}%
			\csname LTb\endcsname
			\put(4793,409){\makebox(0,0){\strut{}$0.9$}}%
		}%
		\gplgaddtomacro\gplfronttext{%
			\csname LTb\endcsname
			\put(153,1904){\rotatebox{-270}{\makebox(0,0){\strut{}$f\sigma_8(z)$}}}%
			\csname LTb\endcsname
			\put(2770,130){\makebox(0,0){\strut{}$z$}}%
			\csname LTb\endcsname
			\put(4005,3046){\makebox(0,0)[r]{\strut{}GLT\ background\ and\ $f\sigma_8$}}%
			\csname LTb\endcsname
			\put(4005,2860){\makebox(0,0)[r]{\strut{}$\Lambda$CDM background and $f\sigma_8$}}%
			\csname LTb\endcsname
			\put(4005,2674){\makebox(0,0)[r]{\strut{}GLT\ background}}%
			\csname LTb\endcsname
			\put(4005,2488){\makebox(0,0)[r]{\strut{}Planck}}%
			\csname LTb\endcsname
			\put(4005,2302){\makebox(0,0)[r]{\strut{}$\Lambda$CDM}}%
			\csname LTb\endcsname
			\begin{tiny}
			\csname LTb\endcsname
			\put(787,1535){\makebox(0,0)[l]{\strut{}6dFGS}}%
			\csname LTb\endcsname
			\put(1197,2650){\makebox(0,0)[l]{\strut{}SDSS-MGS}}%
			\csname LTb\endcsname
			\put(2141,1200){\makebox(0,0)[l]{\strut{}BOSS-LOWZ}}%
			\csname LTb\endcsname
			\put(2900,2210){\makebox(0,0)[l]{\strut{}CMASS}}%
			\csname LTb\endcsname
			\put(1150,1200){\makebox(0,0)[l]{\strut{}SDSS-LRG-200}}%
			\csname LTb\endcsname
			\put(3265,1360){\makebox(0,0)[l]{\strut{}WiggleZ}}%
			\csname LTb\endcsname
			\put(2545,1380){\makebox(0,0)[l]{\strut{}WiggleZ}}%
			\csname LTb\endcsname
			\put(3849,1500){\makebox(0,0)[l]{\strut{}WiggleZ}}%
			\csname LTb\endcsname
			\put(3930,1200){\makebox(0,0)[l]{\strut{}Vipers PDR-2}}%
			\end{tiny}
		}%
		\gplbacktext
		\put(0,0){\includegraphics{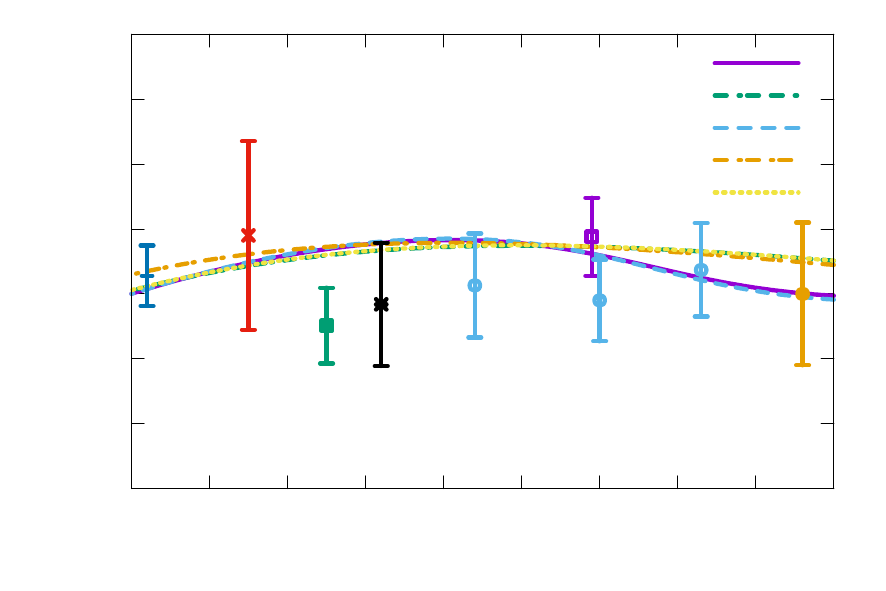}}%
		\gplfronttext
		\end{picture}%
		\endgroup

	\caption{We have plotted the $f\sigma_8(z)$ for the best fit values of GLTofDE. Our model has its own fingerprint which can be seen around $z\sim 0.75$.}
	\label{fig:fs8}
\end{figure}

\end{document}